\documentclass[onecolumn,amsmath,amssymb,aps,showpacs,superscriptaddress,nofootinbib,noshowpacs]{revtex4}

\usepackage[dvips]{graphicx}
\usepackage[dvips]{color}
\usepackage{hyperref,breakurl,amsmath,amssymb,pst-node,eepic}
\usepackage{graphicx}
\usepackage{dcolumn}
\usepackage{bm}
\usepackage{hyperref}
\usepackage{float}
\usepackage{amsmath}
\usepackage{amsfonts}

\usepackage{booktabs}

\begin{document}
	
	\title{When facts fail: Bias, polarisation and truth in social networks}
	
	\author{Orowa Sikder} 
	\affiliation{Department of Computer Science, University College London, Gower Street, London WC1E 6EA, UK}
	\author{Robert E. Smith}
	\affiliation{Department of Computer Science, University College London, Gower Street, London WC1E 6EA, UK}
	\author{Pierpaolo Vivo}
	\affiliation {Department of Mathematics, King's College London, Strand, London WC2R 2LS, UK}
	\author{Giacomo Livan}
	\affiliation{Department of Computer Science, University College London, Gower Street, London WC1E 6EA, UK}
	\affiliation{Systemic Risk Centre, London School of Economics and Political Sciences, Houghton Street, London WC2A 2AE, UK}
	
	\date{\today}
	
	\begin{abstract}
Online social networks provide users with unprecedented opportunities to engage with diverse opinions. At the same time, they enable confirmation bias on large scales by empowering individuals to self-select narratives they want to be exposed to. A precise understanding of such tradeoffs is still largely missing. We introduce a social learning model where most participants in a network update their beliefs unbiasedly based on new information, while a minority of participants reject information that is incongruent with their preexisting beliefs. This simple mechanism generates permanent opinion polarization and cascade dynamics, and accounts for the aforementioned tradeoff between confirmation bias and social connectivity through analytic results. We investigate the model's predictions empirically using US county-level data on the impact of Internet access on the formation of beliefs about global warming. We conclude by discussing policy implications of our model, highlighting the downsides of debunking and suggesting alternative strategies to contrast misinformation.
	\end{abstract}
	
	\maketitle
	
	\section*{Introduction}

We currently live in a paradoxical stage of the information age. The more we gain access to unprecedented amounts of knowledge thanks to digital technologies, the less our societies seem capable of discerning what is true from what is false, even in the presence of overwhelming evidence in support of a particular position. For example, large segments of our societies do not believe in the reality of climate change \cite{mccright2011politicization} or believe in the relationship between vaccinations and autism \cite{horne2015countering}.
	
As recent studies indicate, over two-thirds of US adults get information from online and social media, with the proportion growing annually \cite{levy2014reuters,gottfried2016news}. Hence, the impact such media have in shaping societal narratives cannot be understated. Online media empower their users to choose the news sources they want to be exposed to. This, in turn, makes it easier to restrict exposure only to narratives that are congruent to pre-established viewpoints \cite{mitchell2014political,nikolov2015measuring,conover2011political,schmidt2017anatomy}, and this positive feedback mechanism is further exacerbated by the widespread use of personalized news algorithms \cite{hannak2013measuring}. In other words, \emph{confirmation bias} \cite{jonas2001confirmation,nickerson1998confirmation} is enabled at unprecedented scales \cite{del2016spreading}. 
	
Another major impact of digital technologies has been the increase in connectivity fostered by the growth of online social networks, which plays a double-edged role. On the one hand, it can compound the effects of confirmation bias, as users are likely to re-transmit the same information they are selectively exposed to, leading to fragmented societies that break down into online ``echo chambers'' where the same opinions keep being bounced around \cite{del2016spreading,an2014sharing}. On the other hand, it also translates into a potentially increased heterogeneity of the information and viewpoints users are exposed to \cite{messing2014selective,bakshy2015exposure}.
	
Online social networks can therefore both improve and restrict the diversity of information individuals engage with, and their \emph{net} effect is still very much debated. Empirical research is still in its infancy, with evidence for both positive and negative effects being found \cite{bakshy2015exposure,flaxman2016filter,goel2010real}. The theoretical literature is lagging somewhat further behind. While there exist a plethora of models related to information diffusion and opinion formation in social networks, a sound theoretical framework accounting for the emergence of the phenomena that are relevant to modern information consumption (rather than explicitly introducing them ad hoc), is still largely lacking.
	
In bounded confidence models \cite{hegselmann2002opinion,del2017modeling,quattrociocchi2014opinion} agents only interact with others sharing similar opinions, and thus are characterized by a form of confirmation bias. In such models polarization is a natural outcome assuming agents are narrow enough in their choice of interaction partners \cite{lorenz2007continuous}. However, these models tend to lack behavioural micro-foundations and a clear mechanism to link information diffusion to opinion formation, making it hard to draw conclusions about learning and accuracy amongst agents.
	
Social learning models \cite{demarzo2003persuasion,golub2010naive,acemoglu2011opinion} provide a broader, empirically grounded, and analytically tractable framework to understand information aggregation \cite{mobius2014social}. Their main drawback, however, is that by design they tend to produce long run population consensus, hence fail to account for any form of opinion heterogeneity or polarization \cite{golub2017learning}. Polarization can be generated by introducing ``stubborn'' agents that remain fully attached to their initial opinions rather than interacting and learning from their neighbors \cite{acemouglu2013opinion,acemoglu2010spread}, a mechanism reminiscent of confirmation bias. However, the conditions under which polarization occurs are very strict, as populations converge towards consensus as soon as stubborn agents accept even a negligible fraction of influence from their neighbors \cite{golub2017learning}. A similar phenomenon is explored in social physics literature where it is referred to as networks with ``zealots'', which similarly impede consensus, such as in \cite{redner}. A key distinction in the model we introduce in the following is that all agents are free to vary their opinions over time, resulting in cascade dynamics that separate consensus and polarization regimes.
	
Overall, while it is clear from the literature that some notion of ``bias'' in networks is a key requirement for us to reproduce realistic dynamics of opinion formation, it is still difficult to provide a unified framework that can account for information aggregation, polarization and learning. The purpose of the present paper is to develop a framework that naturally captures the effect of large-scale confirmation bias on social learning, and to examine how it can drastically change the way a networked, decentralized, society processes information. We are able to provide analytic results at all scales of the model. At the macroscopic scale, we determine under what conditions the model ends up in a polarised state or cascades towards a consensus. At the mesoscopic scale, we are able to provide an intuitive chracterization of the trade-off between bias and connectivity in the context of such dynamics, and explain the role echo chambers play in such outcomes. At the microscopic scale, we are able to study the full distribution of each agent's available information and subsequent accuracy, and demonstrate that small amounts of bias can have positive effects on learning by preserving information heterogeneity. Our model unveils a stylized yet rich phenomenology which, as we shall discuss in our final remarks, has substantial correspondence with the available empirical evidence.

\section*{Results}

\subsection*{Social learning and confirmation bias}

	We consider a model of a social network described by a graph $G = (V,E)$ consisting of a set of agents $V$ (where $\vert V \vert$ = $n$), and the edges between them $E$. Each agent seeks to learn the unobservable ground truth about a binary statement $X$ such as, e.g., ``global warming is / is not happening'' or ``gun control does / does not reduces crime''. The value $X = +1$ represents the statement's true value, whose negation is $X=-1$.
	
	Following standard social learning frameworks \cite{mobius2014social}, at the beginning of time ($t = 0$), each agent $i$ ($i = 1, \ldots, n$) independently receives an initial signal $s_i = \pm 1$, which is informative of the underlying state, i.e. $p=\mathrm{Prob}(s_i=+1 | X=+1)=1-\mathrm{Prob}(s_i=-1 | X=+1) >  1/2$. Signals can be thought of as news, stories, quotations, etc., that support or detract from the ground truth. The model evolves in discrete time steps, and at each time step $t > 0$ all agents synchronously share with their neighbors the full set of signals they have accrued up to that point. For example, the time $t=1$ information set of an agent $i$ with two neighbors $j$ and $\ell$ will be $\boldsymbol{s}_i(t=1) = \{s_i,s_j,s_\ell\}$, their time $t=2$ set will be $\boldsymbol{s}_i(t=2) = \{s_i,s_i,s_i,s_j,s_j,s_\ell,s_\ell,\boldsymbol{s}_{d = 2}\}$, where $\boldsymbol{s}_{d = 2}$ denotes the set of all signals incoming from nodes at distance $d = 2$ (i.e., $j$'s and $\ell$'s neighbors), and so on. Furthermore, we define the following:
	
	\begin{equation}
	\label{eq:signal_mix}
	x_i(t) = \frac{N_i^+(t)}{N_i^+(t)+N_i^-(t)} \ ,
	\end{equation} 
	where $N_i^+(t)$ and $N_i^-(t)$ denote, respectively, the number of positive and negative signals accrued by $i$ up to time $t$. We refer to this quantity as an agent's \emph{signal mix}, and we straightforwardly generalize it to any set of agents $C \subseteq V$, i.e., we indicate the fraction of positive signals in their pooled information sets at time $t$ as $x_C(t)$. The list of all agents' signal mixes at time $t$ is vectorised as $\boldsymbol{x}(t)$.
	
	Each agent forms a posterior belief of the likelihood of the ground truth given their information sets using Bayes' rule. This is done under a bounded rationality assumption, as the agents fail to accurately model the statistical dependence between the signals they receive, substituting it with a naive updating rule that assumes all signals in their information sets to be independent ($\mathrm{Prob}(X|\boldsymbol{s}_i(t)) = \mathrm{Prob}(\boldsymbol{s}_i(t)|X)\mathrm{Prob}(X) / \mathrm{Prob}(\textbf{s}_i(t))$, where $\mathrm{Prob}(\boldsymbol{s}_i(t)|X)$ is computed as a factorization over probabilities associated to individual signals, i.e. $\mathrm{Prob}(\boldsymbol{s}_i(t)|X) = \prod_c \mathrm{Prob}\left(\boldsymbol{s}_i^{(c)}(t)|X \right)$, where $\boldsymbol{s}_i^{(c)}(t)$ denotes the $c$th component of the vector), which is a standard assumption of social learning models \cite{mobius2014social}. Under such a framework (and uniform priors), the best guess an agent can make at any time over the statement $X$ given their information set is precisely equal to their orientation $y_i(t)$, where $y_i(t) = +1$ for $x_i(t) > 1/2$ and $y_i(t) = -1$ for $x_i(t)<1/ 2$ (without loss of generality, in the following we shall choose network structures that rule out the possibility of $x_i(t) = 1/2$ taking place). The orientations of all $n$ agents at time $t$ are vectorized as $\boldsymbol{y}(t)$; the fraction of positively oriented agents in a group of nodes $C \subseteq V$ is denoted as $y_C(t)$.
	
	The \emph{polarization} $z_C(t) = \min (y_C(t),1-y_C(t))$ of the group $C$ is then defined as the fraction of agents in that group that have the minority orientation. Note that polarization equals zero when there is full consensus and all agents are either positively or negatively oriented. It is maximized when there are exactly half the group in each orientation.
	
	It is useful to think of $\boldsymbol{x}(t)$, $\boldsymbol{y}(t)$ and $z_V(t)$ as respectively representing the pool of available signals, the conclusions agents draw on the basis of the available signals, and a summary measure of the heterogeneity of agents' conclusions. In the context of news diffusion, for example, they would represent the availability of news of each type across agents, the resulting agents' opinions on some topic, and the extent to which those opinions have converged to a consensus.
	
	We distinguish between two kinds of agents in the model: \textit{unbiased} agents and \textit{biased} agents. Both agents share signals and update their posterior beliefs through Bayes' rule, as described in the previous section. However, they differ in how they acquire incoming signals. Unbiased agents accept the set of signals provided by their neighbours without any distortion. On the other hand, biased agents exercise a model of \textit{confirmation bias} \cite{kunda1990case,nickerson1998confirmation}, and are able to distort the information sets they accrue. We denote the two sets of agents as $\mathcal{U}$ and $\mathcal{B}$, respectively.
	
	To describe the behaviour of these biased agents we use a slight variation of the confirmation bias model introduced by Rabin and Shrag \cite{rabin1999first}. We refer to an incoming signal $s$ as \textit{congruent} to $i$ if it is aligned with $i$'s current orientation, i.e. if $s = y_i(t)$, and \textit{incongruent} if $s = - y_i(t)$. When biased agents are presented with incongruent signals, they reject them with a fixed probability $q$ and replace them with a congruent signal, which they add to their information set and propagate to their neighbors. We refer to $q$ as the \textit{confirmation bias} parameter. Denote the set of positively (negatively) oriented biased agents at time $t$ as ${\mathcal{B}^+}(t)$ (${\mathcal{B}^-}(t)$), and the corresponding fraction as $y_{\mathcal{B}}(t) = \vert {\mathcal{B}^+}(t) \vert / \vert \mathcal{B} \vert$. Note that this is an important departure from ``stubborn agent'' models, as such biased agents do have a non-zero influence from their neighbours, and they can change their beliefs over time as they aggregate information.
	
\begin{figure}[h]
		\centering
		\includegraphics[scale=0.4]{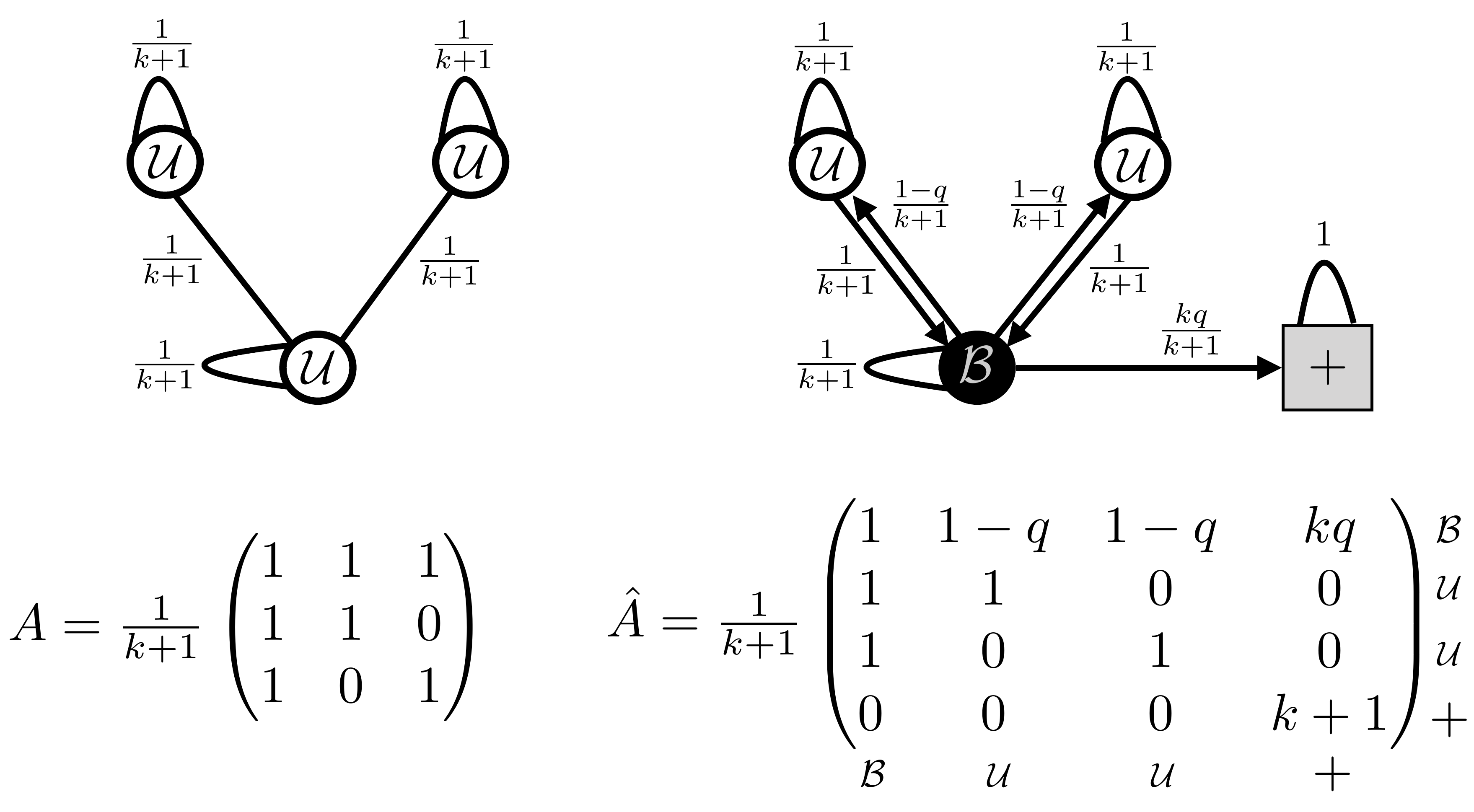}
		\caption{Sketch of an unbiased (left) and biased (right) network with three nodes. Below each sketch is the update matrix of the corresponding (deterministic) DeGroot model. The network on the right shows that the stochastic ``signal distortion'' behaviour can be approximated with a biased agent ($\mathcal{B}$) reducing the weight it places on each of its neighbors to $(1-q)/(k+1)$, and placing the remaining weight $kq/(k+1)$ on an external positively oriented ``ghost'' node. When a biased agents changes orientation, it switches such an edge to an external negatively oriented node instead.}
		\label{fig:DeGroot}
	\end{figure}

	An intuitive interpretation of what this mechanism is intended to model is as follows: biased agents are empowered to reject incoming signals they disagree with, and instead refer to preferred sources of information to find signals that are congruent with their existing viewpoint (see Fig. \ref{fig:DeGroot}). This mechanism models both \emph{active} behaviour, where agents deliberately choose to ignore or contort information that contradicts their beliefs (mirroring the ``backfire effect'' evidenced both in psychological experiments \cite{redlawsk2002hot,nyhan2010corrections} and in online social network behaviour \cite{zollo2017debunking}), and \emph{passive} behaviour, where personalized news algorithms filter out incongruent information and select other information which coheres with the agents' beliefs \cite{bakshy2015exposure}.
	
	In the following, we shall denote the fraction of biased agents in a network as $f$. We shall refer to networks where $f=0$ as \textit{unbiased} networks, and to networks where $f > 0$ as \emph{biased} networks.

	For the bulk of the analytic results in the paper, we assume that the social network $G$ is an undirected $k$-regular network. The motivation for this is two-fold. Firstly, empirical research \cite{bessi2015viral} suggests that for online social networks such as Facebook (where social connections are symmetric), heterogenous network features such as hubs do not play a disproportionately significant role in the diffusion of information. Intuitively, while social networks themselves might be highly heterogenous, the network of information transmission is a lot more restricted, as individuals tend to discuss topics with a small group such as immediate friends and family. Secondly, utilizing a simple $k$-regular network allows for considerable analytical tractability. However, one can show that our main results can be easily extended to hold under a variety of network topologies characterized by degree heterogeneity.
    	
    The assumption of regular network structure (coupled with the aforementioned synchronous belief update dynamics) allows the information sets of all agents to grow at the same rate, and as a result the evolution of the signal mix $\boldsymbol{x}(t)$ can be mapped to a DeGroot averaging process for unbiased networks \cite{degroot1974reaching}: $\boldsymbol{x}(t) = A \ \boldsymbol{x}(t-1)$, 	where $A$ is an $n \times n$ matrix with entries $a_{ij} = 1/(k+1)$ for each pair $(i,j)$ of connected nodes.
	
	For biased networks, one can demonstrate (see Section S1 of the Supplementary Materials) that the above confirmation bias mechanics can be reproduced by introducing a positive and negative ``ghost'' node which maintain respective signal mixes of $1$ and $0$. Biased agents sample each signal from their orientation-aligned ghost node nodes with probability $q$, and from their neighbourhood with probability $(1-q)$.
	
	Furthermore, while the process is stochastic, we also show that it converges to a deterministic process with a simple update matrix described as follows. In Section S3 of the Supplementary Materials we discuss in detail the correspondence and convergence between the stochastic and deterministic processes. Biased agents down-weight their connections to neighbours by a factor $(1-q)$ and place the remaining fraction $kq/(k+1)$ of their outgoing weight on the corresponding ghost node. With these positions the updating process now simply reads $\boldsymbol{\hat{x}}(t) = \hat{A}(t) \boldsymbol{\hat{x}}(t-1)$, where $\hat{A}(t)$ is an $(n+2) \times (n+2)$ asymmetric matrix whose entries in its $n \times n$ upper-left block are as those in $A$, except $\hat{a}_{ij} = (1-q)/(k+1)$ when $i \in \mathcal{B} = \mathcal{B}^+ \cup \mathcal{B}^-$. The positive (negative) ghost node corresponds to node $n+1$ ($n+2$) of the augmented matrix $\hat{A}(t)$, and we shall label it as $+$ ($-$) for convenience, i.e., we shall have $\hat{a}_{i+} = kq/(k+1)$ ($\hat{a}_{i-} = kq/(k+1)$) for $i \in \mathcal{B}^+$ ($i \in \mathcal{B}^-$) and $\hat{a}_{++} = \hat{a}_{--} = 1$. Similarly, $\boldsymbol{\hat{x}}(t)$ denotes an augmented signal mix vector where $\hat{x}_+(t) = \hat{x}_{n+1}(t) = 1$, and $\hat{x}_-(t) = \hat{x}_{n+2}(t) = 0$. 
	
	The time dependence of the matrix $\hat{A}(t)$ is due to the fact that whenever a biased agent switches orientation its links to the ghost nodes change. This happens whenever the agent's signal mix $x_i(t)$ (see Eq. \ref{eq:signal_mix}) goes from below to above $1/2$ or vice versa, due to an overwhelming amount of incongruent incoming signals from its neighbors. For example, when switching from being positively to negatively oriented, a biased agent $i$ will change its links as follows: $\hat{a}_{i+} = kq/(k+1) \rightarrow \hat{a}_{i+} = 0$, and $\hat{a}_{i-} = 0 \rightarrow \hat{a}_{i-} = kq/(k+1)$. In the following, all quantities pertaining to biased networks will be denoted with a $\hat{}$ symbol. 
	
	We provide a sketch of the above mapping in Fig. \ref{fig:DeGroot}. There is an appealing intuition to this interpretation: biased agents have a ``preferred'' information source they sample from in lieu of incongruent information provided from their peers. If their beliefs change, their preferred information source can change.
	
	Our main focus will be on the long-run properties of the dynamics introduced above. In this respect, it is important to establish whether the agents actually reach an equilibrium over their signal mixes and orientations, or whether they continue to oscillate. It is straightforward to demonstrate that unbiased networks always converge to a limiting steady state for their signal mixes and orientations, which follows directly from the correspondence between such networks and DeGroot models \cite{golub2010naive}. The convergence of biased networks is much less trivial to prove due to the non-linear dynamics introduced by the confirmation bias mechanics. It can be shown however that convergence holds under fairly general conditions, and in Section S2 of the Supplementary Materials we demonstrate such convergence under a number of network topologies.
	
	Building on this, we are able to establish the following results on $k$-regular networks of any size (in the following, and throughout the rest of the paper, we shall denote the steady state value of a variable $v$ as $v^*$).
	
    For biased $k$-regular networks, denote $t^*$ as the time after which biased agents cease switching their orientation. Define $\hat{y}_{\mathcal{B}}^*$ as the steady state fraction of positively oriented biased agents. Then the following holds (see Section S3 of the Supplementary Materials).
	
	\begin{itemize}
	
    \item[(1)] The signal mix vector $\boldsymbol{\hat{x}}(t)$ converges to some $\boldsymbol{\hat{x}}^* = \hat{A}^*\boldsymbol{\hat{x}}(0)$ for both biased and unbiased networks, where $\hat{A}^*$ is a steady-state matrix of influence weights which can be computed explicitly (see Section S3 of the Supplementary Materials).
	
    \item[(2)] Unbiased networks achieve consensus, and converge to influence weights of $a^*_{ij} = 1/n$ for all pairs $(i,j)$. This ensures that, for all $i \in V$, $x^*_i=x^*_V=\bar{x}(0)$, where $\bar{x}(0) = \sum_{i=1}^n s_i$ is the intial average signal mix.
	
    \item[(3)] Biased networks where $\hat{y}_{\mathcal{B}}^* = 0,1$ achieve consensus, and converge to influence weights $\hat{a}^*_{ij} = 0$ for all pairs $(i, j) \in V$, $\hat{a}^*_{i+} = \hat{y}_{\mathcal{B}}^*$ and $\hat{a}^*_{i-} = 1-\hat{y}_{\mathcal{B}}^*$ for all $i \in V$.
	
    \item[(4)] Biased networks where $ 0 < \hat{y}_{\mathcal{B}}^* < 1$ do not achieve consensus, and converge to influence weights $\hat{a}^*_{ij} = 0$ for all $(i, j) \in V$, and $\hat{a}^*_{i+} + \hat{a}^*_{i-}=1$ for all $i \in V$.
    \end{itemize}    
	
	From the above, we can conclude that while unbiased networks efficiently aggregate the information available to them at $t=0$, the outcome of the information aggregation process in biased networks ends up being \emph{entirely} determined by the long-run orientations of biased agents. We shall devote the following sections to examine the consequences of the model in greater detail through mean field approximations coupled with numerical verifications on finite networks.
	
\subsection*{Cascades and consensus}
	
	We begin by studying the signal mix of unbiased agents in biased networks ($\hat{x}_\mathcal{U}^*$) to provide a like for like comparison with the fully unbiased networks. In the context of the diffusion of news, the global signal mix can be thought of as a model of the long term balance of news of different types that survive following the diffusion dynamics.
	
	For unbiased networks, it is demonstrated in \cite{livan2013leaders} that $\hat{x}_\mathcal{U}^* = \bar{x}(0)$. That is, the steady state signal mix in unbiased networks precisely reflects the original, unbiased informative signals injected into the network. Determining the steady state signal mix of biased networks entails considering the interactions between three subpopulations - the unbiased agents $\mathcal{U}$, positively biased agents ${\mathcal{B}^+}$, and negatively biased agents $\mathcal{B}^-$. One can show (see Section S4 of the Supplementary Materials) that this can be approximated as:

	\begin{equation} \label{GSSSM}
	\begin{pmatrix}
	\hat{x}_\mathcal{U}^* \\
	\hat{x}_{{\mathcal{B}^+}}^* \\
	\hat{x}_{\mathcal{B}^-}^*
	\end{pmatrix}
    =
	\begin{pmatrix}
	\hat{y}_{\mathcal{B}}(t^*) \\
	(1-q)\hat{y}_{\mathcal{B}}(t^*) +q \\
	(1-q)\hat{y}_{\mathcal{B}}(t^*)
	\end{pmatrix} \ .
	\end{equation}
	
	Let us now consider the situation under which $t^* = 0$, i.e. where the initial orientation of each biased agent does not change, and is therefore equal to the initial signal it receives. Intuitively, this will occur for large $q$ which allows for biased agents to reject the majority of incongruent signals they receive (shortly we demonstrate in fact this generally occurs for $q>1/2$). Given Eq. \ref{GSSSM}, we can therefore calculate the steady state signal mix of any subset of agents based on our knowledge of the distribution of the initial signals.
	
	In this scenario, the average signal mix $\hat{x}_\mathcal{U}^*$ of unbiased agents is determined by the initial proportion of positively oriented biased agents $y_{\mathcal{B}^+}(0)$, which is the mean of $fn$ i.i.d. Bernoulli variables with probability $p$ (which, we recall, denotes the probability of an initially assigned signal being informative). One can compare this to unbiased networks ($f=0$), where the long run average signal mix is $x_V^* = \bar{x}(0)$, and is hence the mean of $n$ i.i.d. Bernoulli variables with probability $p$. Applying the central limit theorem we see that injecting a fraction $f$ of biased agents therefore amplifies the variance of the long run global signal mix by a factor of $f^{-1}$ with respect to the unbiased case:
	
	\begin{equation} \label{eq:distr_mean_mix}
	x_V^* \sim \mathcal{N} \left (p,\frac{p(1-p)}{n} \right ) \ \rightarrow \ \hat{x}_\mathcal{U}^* \sim \mathcal{N} \left (p,\frac{p(1-p)}{fn} \right) \ .
	\end{equation}
	This means that the ``wisdom of unbiased crowds'' is effectively undone by small biased populations, and the unbiased network's variability is recovered for $f \rightarrow 1$, and not for $f \rightarrow 0^+$, as one might intuitively expect. 
	
	Consider now the general case where biased agents can, in principle, switch orientation a few times before settling on their steady state orientation. Using mean-field methods one can determine the general conditions under which a cascade in these orientation changes can be expected (see Section S4 of the Supplementary Materials) but here we only provide some intuition. As $q$ is lower, it is easier for an initial majority camp of biased agents to convert the minority camp of biased agents. As the conversion of the minority camp begins, this triggers a domino effect as newly converted biased agents add to the critical mass of the majority camp and are able to overwhelm the minority orientation.
	
	This mechanism allows us to derive analytic curves in the parameter space to approximate the steady state outcome of the unbiased agent population's average signal mix based on the orientations of the biased agents at time $0$: 
	
	\begin{equation} \label{eq:curtainplot}
	\hat{x}^*_\mathcal{U} = \left\{
	\begin{array}{ll}
	\hat{y}_{\mathcal{B}}(0) & \text{for} \ \frac{1-2q}{2(1-q)} \leq \hat{y}_{\mathcal{B}}(0) \leq \frac{1}{2(1-q)} \\
	1 & \text{for} \ \hat{y}_{\mathcal{B}}(0) > \frac{1-2q}{2(1-q)} \\
	0 & \text{for} \  \hat{y}_{\mathcal{B}}(0) < \frac{1}{2(1-q)} . \\ 
	\end{array}
	\right.
	\end{equation}
	
\begin{figure}
		\centering
		\includegraphics[scale=0.6]{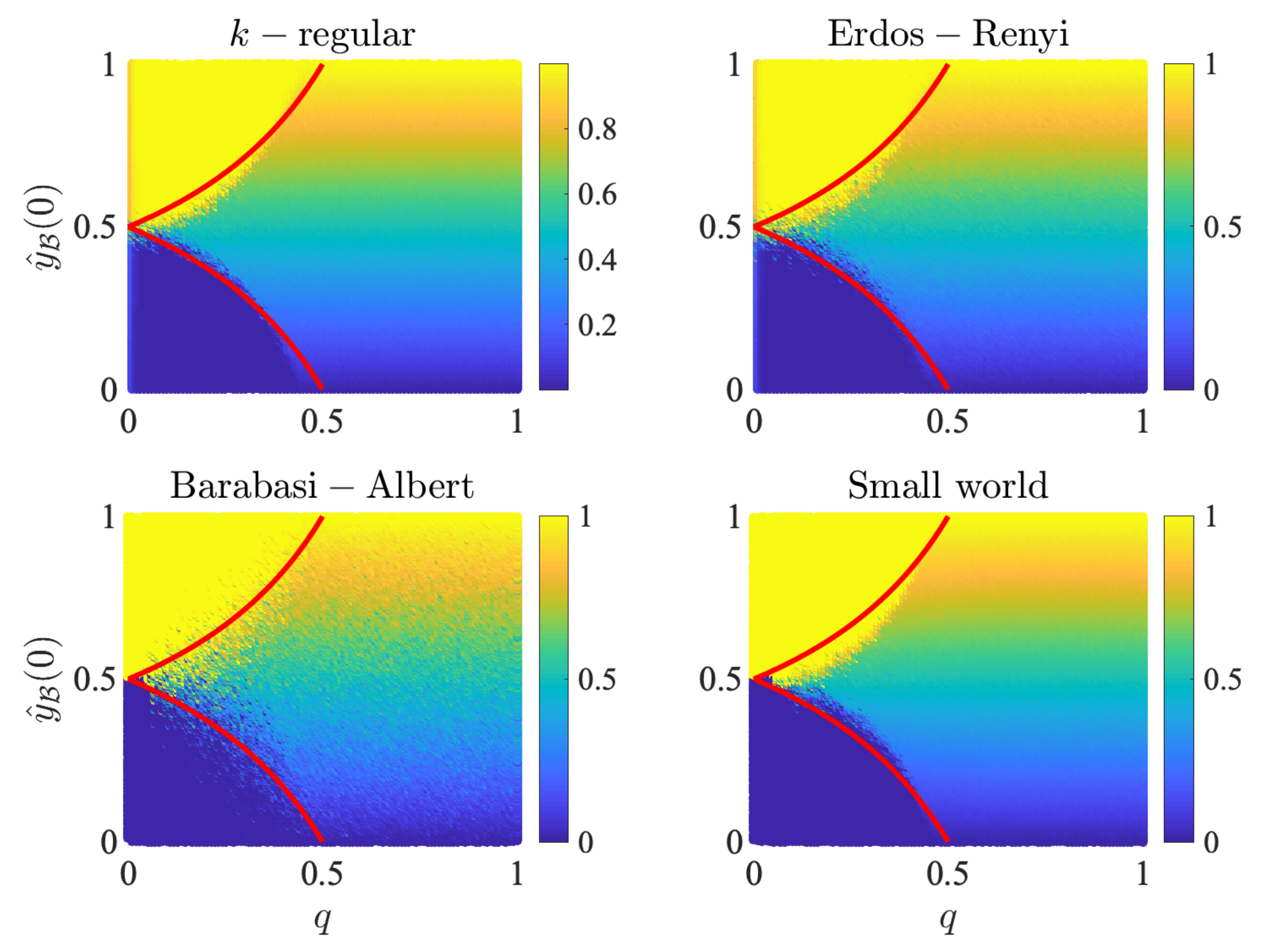}
		\caption{Average steady state signal mix of unbiased agents ($\hat{x}_{\mathcal{U}}^*$) as a function of the time $t=0$ fraction of positively oriented biased agents ($\hat{y}_\mathcal{B}(0)$) and confirmation bias $q$. The color gradient denotes the average long run signal mix for unbiased agents from $0$ to $1$. The top-left and bottom-left regions are characterized, respectively, by a global signal mix of 1 and 0 respectively, and are separated by a discontinuous transition from a region characterized by a steady state that maintains a mixed set of signals. That is, in the top-left region almost all negative signals have been removed from the network, leaving almost entirely positive signals in circulation (and vice versa for the bottom-left region). In the remaining region, signal mixes of both types survive in the long run, and the balance between positive and negative signals reflects the fraction of positively oriented biased agents. The lower $q$ falls, the easier it is to tip the network into a total assimilation of a single signal type. Results are shown for simulations on  $k$-regular (top left), Erd\H os-R\'enyi (top right), Barabasi-Albert (bottom left) and Small-world (bottom right) networks. Analytic predictions (given by Eq. \eqref{eq:curtainplot}) are denoted by solid red lines. The parameters used in the simulations were $n = 10^3$, $p = 0.51$, $k = 6$ (which corresponds to an average degree in all cases), $f = 0.4$.}
		\label{fig:curtainplot}
	\end{figure}

	The above result is sketched in Fig. \ref{fig:curtainplot}, and we have verified that it matches numerical simulations even for heterogenous networks. For $1 / 2 < q \leq 1$ biased agents can convert at least half of the incongruent signals they receive to their preferred type, meaning that biased agents of either orientation cannot be eradicated from the network, which preserves signals of  both types in the steady state. For sufficiently small values of $q$, on the other hand, small variations in the initial biased population translate to completely opposite consensus, and only by increasing the confirmation bias $q$, paradoxically, the model tends back to a balance of signals that resembles the initially available information.
	
	Putting the above results together, we note that biased networks with small $f$ and $q$ are, surprisingly, the most unstable. Indeed, such networks sit on a knife-edge between two extremes where one signal type flourishes and the other is totally censored. In this context, the model indicates that confirmation bias helps preserve a degree of information heterogeneity, which, in turn, ensures that alternative viewpoints and information are not eradicated. In subsequent sections we consider a normative interpretation of this effect in the context of accuracy and learning.

    \subsection*{Polarization, echo chambers and the bias-connectivity trade-off}

	So far we have derived the statistical properties of the average steady state signal mix across \textit{all} unbiased agents. We now aim to establish how these signals are distributed \textit{across} individual agents. Throughout the following, assume the global steady state signal mix $\hat{x}_\mathcal{U}^*$ has been determined.
	
\begin{figure}[tbhp]
		\centering
		\includegraphics[scale=0.65]{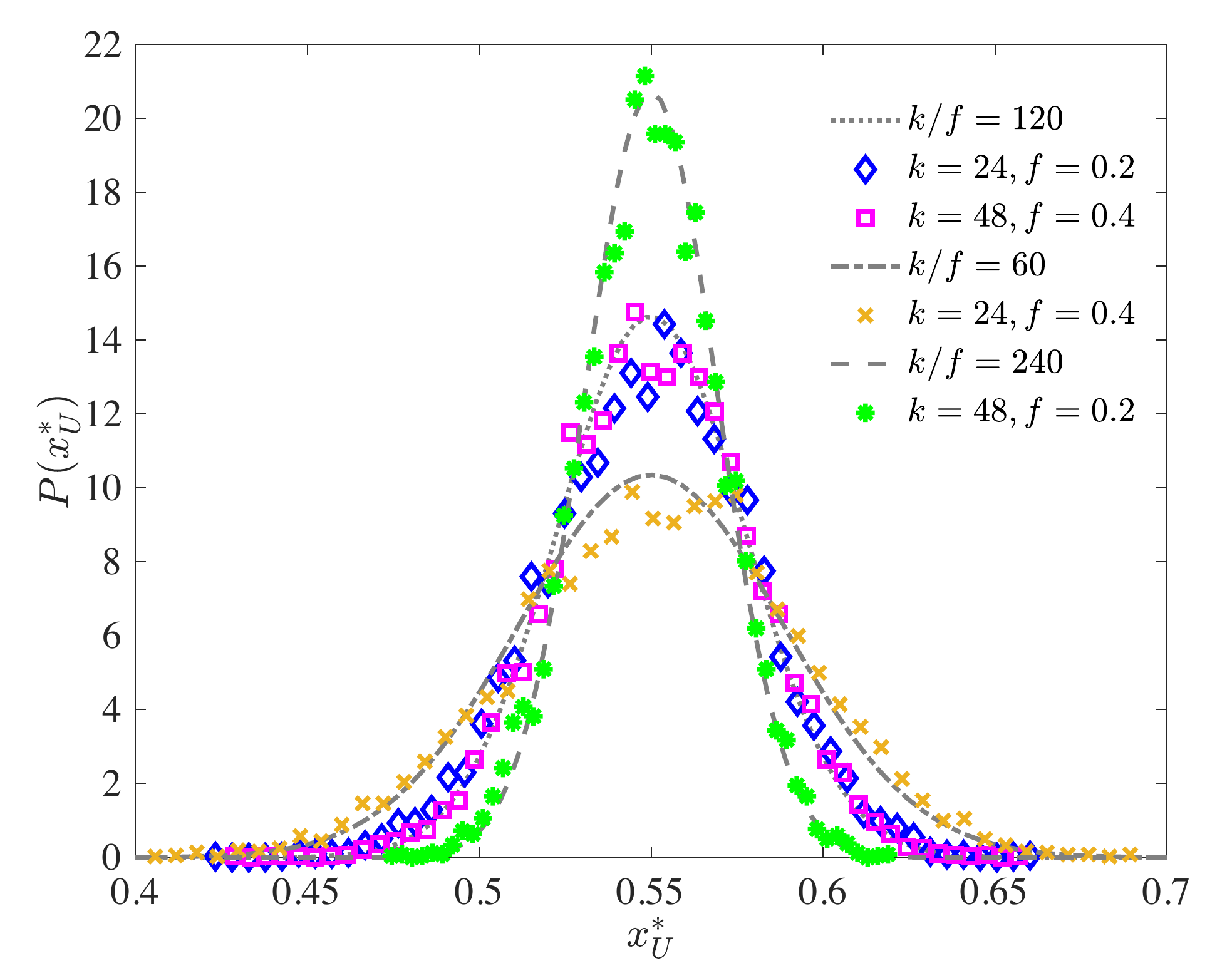}
		\caption{Distribution of individual unbiased agents' steady state signal mixes (points) vs analytic predictions (dashed lines). As discussed in the main text, the model predicts such distribution to be a Gaussian (for both large $n$ and $k$) with mean equal to $\hat{x}^*_{\mathcal{U}}$ and variance given by Eq. (5). $\hat{x}^*_{\mathcal{U}}$ is kept fixed to demonstrate the effect of varying $f$ and $k$. As shown in the case for $k/f = 120$, $f$ and $k$ trade off, and scaling both by the same constant results in the same distribution. The parameters used in the simulations were $n = 10^4$, $\hat{x}^*_{\mathcal{U}} = 0.55$, $q = 0.6$.}
		\label{fig:fktrade}
	\end{figure}

	In the limit of large $n$ and $k$, $x_i^*$ for $i \in \mathcal{U}$ is normally distributed with mean $\hat{x}_\mathcal{U}^*$ and variance $ \sigma^2(\hat{x}_\mathcal{U}^*)$ that can be approximated as follows (see Section S4 of the Supplementary Materials):
	
	\begin{equation} \label{eq:variance}
	\sigma^2(\hat{x}_\mathcal{U}^*) \approx \frac{f q^2}{k} (\hat{x}^*_\mathcal{U}(1-\hat{x}^*_\mathcal{U})) \ ,
	\end{equation}
	and this result is quite accurate even when compared with simulations for small $n$ and $k$, as demonstrated in Figure \ref{fig:fktrade}. 
	
	This result further shows that the presence of biased agents is effectively responsible for the polarization of unbiased agents in the steady state. Indeed, both a larger biased population and higher confirmation bias - i.e. higher $f$ or $q$, respectively - result in an increased variance and steady state polarization $\hat{z}_\mathcal{U}^*$, since a larger variance $\sigma^2(x_\mathcal{U}^*)$ implies larger numbers of agents displaying the minority orientation. This is illustrated in the top left panel of Fig. \ref{fig:accuracy}.

\begin{figure}
		\centering
		\includegraphics[scale=0.4]{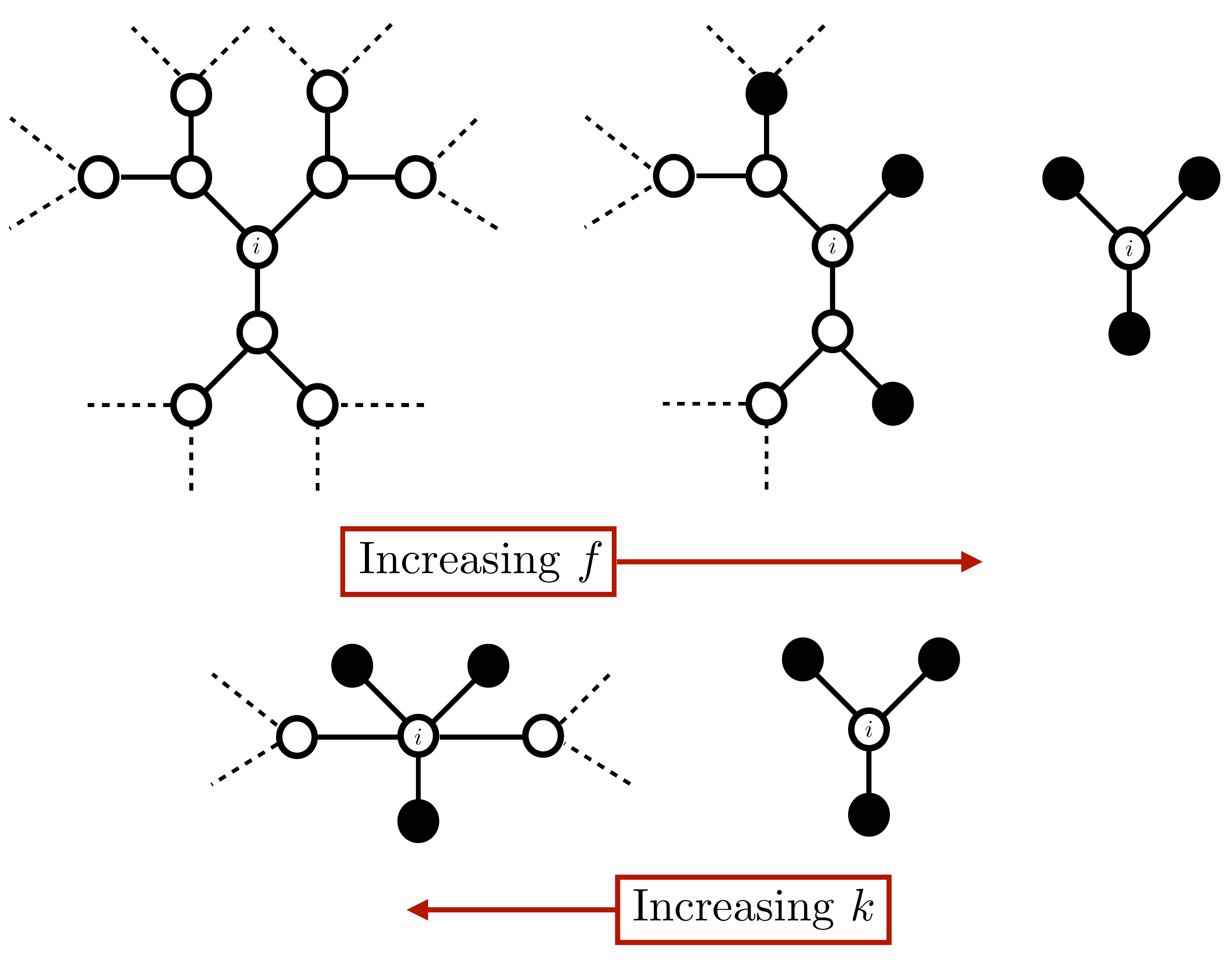}
		\caption{For illustration of the bias-connectivity trade-off, consider a Cayley $3$-tree structure and let $q=1$. When $f=0$, each unbiased node has 3 independent sources of novel information, one from each branch associated with a neighbour. As $f$ increases, unbiased neighbours are steadily replaced with biased agents that act as gatekeepers and restrict the flow of novel information from their branches, resulting in an equivalence with fewer branches overall. Increasing social connectivity $k$ bypasses biased neighbours and enables the discovery of novel information. This can also be interpreted as agents being encased in ``echo chambers'' as bias grows, and circumventing these chambers as connectivity increases.}
		\label{fig:fkintuition}
	\end{figure}

	On the other hand, a larger degree $k$ contrasts this effect by creating more paths to transport unbiased information. It is worth pointing out that the variance in Eq. \ref{eq:variance} does not decay with $n$, showing that steady state polarization persists even in the large $n$ limit. We refer to this as the bias-connectivity trade-off, and the intuition behind this result is illustrated in Figure \ref{fig:fkintuition}.
	
    Further intuition for this result can be found at the mesoscopic level of agent clusters, where we see the emergence of natural ``echo chambers'' in the model. We define an echo chamber $C$ as a subset of unbiased agents such that: $C = \{i \in \mathcal{U}: \partial_i \in \mathcal{B}\cup C\ \land \partial_i \cap C \neq \emptyset\}$, where $\partial_i$ denotes the neighbourhood of agent $i$. In other words, an echo chamber is a set of connected unbiased agents such that all nodes are either connected to other nodes in the echo chamber or to biased agents. Therefore, biased agents form the echo chamber's boundary, which we refer to as $\partial_C$. Echo chambers in our model represent groups of unbiased agents that are completely surrounded by biased agents who effectively modulate the information that can flow in and out of these groups.
    
    Echo chambers allow us to examine the qualitative effect of confirmation bias ($f, q$) and connectivity $k$. Let us label the fraction of unbiased agents enclosed in an echo chamber as $\eta_C$. Leveraging some simple results from percolation theory\cite{perc1} we can show that $\eta_C$ increases with $f$ and decreases with $k$, as the creation of more pathways that bypass biased agents effectively breaks up echo chambers. Furthermore, the equilibrium signal mix of unbiased agents inside echo chambers is well approximated by a weighted average between the signal mix of the biased agents surrounding them ($x^*_{\partial_C}$) and the signal mix $x_{\mathcal{U}}^*$ of the whole population: $x_C^* = q \ x^*_{\partial C} + (1-q) x_{\mathcal{U}}^*$. The confirmation bias parameter $q$ therefore determines the ``permeability'' of echo chambers to the information flow from the broader network. Hence, unbiased agents enclosed in echo chambers are likely to be exceedingly affected by the views of the small set of biased agents surrounding them, and, as such, to hold information sets that are unrepresentative of the information available to the broader network. In doing so, we can envision these echo chambers as effective ``building blocks'' of the overall polarization observed in the network.
    

\subsection*{Accuracy, efficiency and learning}
	
	Up until now, we have not attempted to make any normative interpretations of the ground truth $X=+1$. In the following, we shall refer to unbiased agents whose steady state orientation is positive (negative) as accurate (inaccurate) agents, and we shall define the overall accuracy $\mathcal{A}(G)$ of a network $G$ as the \textit{expected} fraction of accurate agents in the steady state. This allows us to investigate how biased and unbiased networks respond to changes in the reliability of the available information, which ultimately depends on the prevalence of positive or negative signals (modulated by the parameter $p = \mathrm{Prob} ( s = +1 | X = +1 )$), which, loosely speaking, can be interpreted as ``real'' and ``fake'' news.
	
	The accuracy of unbiased networks obtains a neat closed form that can be approximated as $\mathcal{A}(G | f=0) \approx \mathrm{erfc} ( (1-2p) \sqrt{n / 2} ) / 2$ (see \cite{livan2013leaders}). For $f>0$, we compute the expected accuracy as the expected fraction of accurate agents with respect to a certain global signal mix. This reads:
	
	\begin{equation} \label{eq:accuracy}
	\mathcal{A}(G | f>0) = \frac{1}{2} \int_0^1 \mathrm{d} x_\mathcal{U}^* \ P(x_\mathcal{U}^*) \ \mathrm{erfc} \left (\frac{1/2 - x_\mathcal{U}^*}{\sqrt{2}\sigma_{x_\mathcal{U}^*}} \right ) \ ,
	\end{equation}
	where $P(x_\mathcal{U}^*)$ is the distribution of the average signal mix across unbiased agents (see Eq. \ref{eq:distr_mean_mix}) (we take the simplifying case of $q > 1/2$, but this can easily be extended to the case for $q \leq 1/2$ using Eq. \ref{eq:curtainplot}), and where we have used the previously mentioned Gaussian approximation for the distribution of individual signal mixes (whose variance $\sigma_{x_\mathcal{U}^*}^2$ is given by  Eq. \ref{eq:variance}).
	
	\begin{figure}
		\centering
		\includegraphics[scale=0.7]{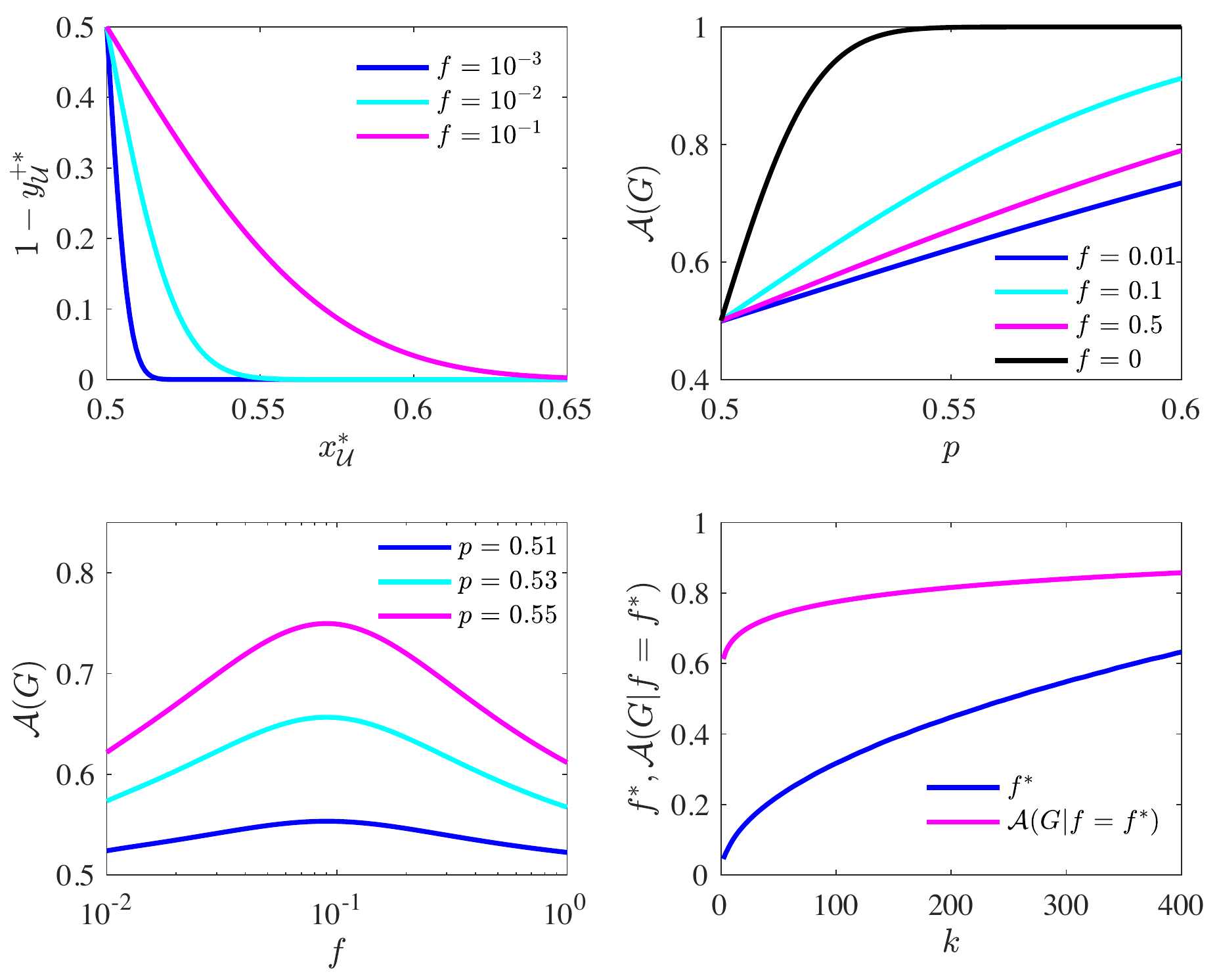}
		\caption{polarization $\hat{z}_\mathcal{U}^* = 1-\hat{y}^{+*}_\mathcal{U}$ of the unbiased agent population as a function of the average signal mix $\hat{x}^*_\mathcal{U}$ in the steady state calculated as $\mathrm{erfc} (\hat{x}^*_\mathcal{U} - 1/2) / (\sqrt{2}\sigma(\hat{x}^*_\mathcal{U})))/2$,  with $\sigma(\hat{x}^*_\mathcal{U})$ given by Eq. \ref{eq:variance} (top left panel). Expected accuracy (Eq. \ref{eq:accuracy}) as a function of the initial signals' informativeness $p$ (top right panel) and of the fraction $f$ of biased agents (bottom left panel). behaviour of the accuracy-maximizing value $f^*$ and of the corresponding accuracy $\mathcal{A}(G | f = f^*)$ as functions of $k$ (bottom right panel). In the first three panels the model's parameter $n=10^3$, $q = 1$, $k=8$, while the parameters in the last panel are $n=10^3$, $q = 1$, $p = 0.53$. In all cases we assume $X=+1$ without loss of generality.}
		\label{fig:accuracy}
	\end{figure} 
	
	The top right panel in Fig. \ref{fig:accuracy} contrasts biased and unbiased networks, and shows how the former remain very inefficient in aggregating information compared to the latter, even as the reliability of the signals ($p$) improve. However, accuracy in biased networks is non-monotonic with respect to $f$. As shown in the bottom left panel in Fig. \ref{fig:accuracy}, accuracy reaches a maximum in correspondence of an optimal value $f^*$ (see Section S5 of the Supplementary Materials for a comparison with numerical simulations). Intuitively, this is because for small values of $f$, as already discussed, the model can converge to the very inaccurate views of a small set of biased agents. As $f$ grows, the views of the two biased camps tend to cancel each other out, and the signal set will match more closely the balance of the original distribution of signals (Eq. \ref{eq:distr_mean_mix}). However, in doing so large values of $f$ lead to increased polarization (Eq. \ref{eq:variance}), where accurate and inaccurate agents coexist. The trade-off between balance and polarization is optimised at $f^*$.
	
	It is also interesting to note that, as shown in the bottom right panel of Fig. \ref{fig:accuracy}, the optimal fraction of biased agents $f^*$ and the corresponding maximum accuracy $\mathcal{A}(G | f = f^*)$ both increase monotonically with the degree. This indicates that as networks are better connected, they can absorb a greater degree of confirmation bias without affecting accuracy.
	
\subsection*{Internet access, confirmation bias, and social learning}
	
    We now seek to test some of the model's predictions against real world data. Clearly, a full validation of the model will require an experimental setup, but a simple test case on existing data can demonstrate the utility of the framework in disambiguating the competing effects of bias and connectivity. We employ the model to investigate the effect of online media in the process of opinion formation using survey data.  Empirical literature on this phenomenon has been mixed, with different analyses reaching completely opposite conclusions, e.g., showing that Internet access increases \cite{lelkes2017hostile}, decreases \cite{barbera2014social} and has no effect \cite{boxell2017greater} on opinion polarization. In Section S6 of the Supplementary Materials we briefly review how our model can help better understand some of the inconsistencies between these results.
	
	Our position is that the effect of Internet access can be split into the effect it has on social connectivity and social discussion ($k$) and the residual effect it has on enabling active and passive confirmation bias behaviours ($f$). As per Eq. \ref{eq:variance}, assuming the majority of the population accurately learns the ground truth ($x_\mathcal{U}^*>1/2$), increases in social discussion should improve consensus around the truth and reduce the fraction of inaccurate agents. However, when controlling for the improvement in social connectivity, we should expect an increase in Internet access to have the opposite effect.
	
	We utilise data from the Yale Programme on Climate Change Communication \cite{howe2015geographic}, which provides state and county level survey data on opinions on global warming, as well as information about the propensity to discuss climate change with friends and family, which proxies connectivity $k$. We combine this with FCC reports on county level broadband internet penetration, which proxies for $f$ after controlling for the considerable effect this has on social connectivity. We also account for a range of covariates (income, age, education, etc) and make use of an instrumental variable approach to account for simultaneous causality. We then attempt to predict the fraction of each county's population that correctly learns that ``global warming is happening'' (see Section S6 of the Supplementary Materials for details on assumptions and results).
	
	As predicted by our model, we find the accurate fraction of the population to have statistically significant positive relationships with $k$, and a negative relationship with $f$. We find such relationships to account for $65\%$ of the variance in the data. This indicates that, after controlling for the improvements on social connectivity, Internet access does indeed increase polarization and reduces a population's ability to accurately learn. While simple, this analysis illustrates the value of our model: by explicitly accounting for the separate effects of large-scale online communication (confirmation bias and connectivity), it can shed light on the mixed empirical results currently available in the literature. In Section S6 of the Supplementary Materials we explore this further by reviewing some of these empirical results and showing how our model provides useful further interpretations of available findings.
	
    It should be emphasized that this result is merely an initial exploration of how our model can provide some testable predictions to empirical data, as opposed to a detailed effort to understand the effect of Internet access on global warming beliefs. Having said that, the initial results are encouraging, and we hope the clarity of the analytic results of our model pave the way for testing variations of the idea of biased information aggregation in a range of outcomes and settings.

\section*{Discussion}

	We introduced a model of social learning in networked societies where only a fraction of the agents update beliefs unbiasedly based on the arrival of new information. The model only provides a stylized representation of the real-world complexity underpinning the propagation of information and the ensuing opinion formation process. Its value stands in the transparency of the assumptions made, and in the fact that it allows us to ``unpack'' blanket terms such as, e.g., social media and Internet penetration, by assigning specific parameters to their different facets, such as connectivity ($k$) and the level of confirmation bias it enables in a society ($f, q$). This, in turn, yields quantitative \emph{testable} predictions that contribute to shed light on the mixed results that the empirical literature has so far collected on the effects online media have in shaping societal debates.
	
	Our model indicates the possibility that the ``narratives'' (information sets) biased societies generate can be \emph{entirely} determined by the composition of their sub-populations of biased reasoners. This is reminiscent of the over-representation in public discourse of issues that are often supported by small but dedicated minorities, such as GMO opposition \cite{blancke2015fatal}, and of the domination of political news sharing on Facebook by heavily partisan users \cite{bakshy2015exposure}; it also resonates with recent experimental results showing that committed minorities can overturn established social conventions \cite{centola2018experimental}. The model indicates that societies that contain only small minorities of biased individuals ($f \rightarrow 0^+$) may be much more prone to producing long run narratives that deviate significantly from their initially available information set (see Eq. \ref{eq:distr_mean_mix}) than societies where the vast majority of the agents actively propagate biases. This resonates, for example, with Gallup survey data about vaccine beliefs in the US population, where only $6\%$ of respondents report their belief in the relationship between vaccines and autism, but more than $50\%$ report to be unsure about it and almost $75\%$ report to have heard about the disadvantages of vaccinations \cite{newport2015us}. Similarly, the model suggests that mild levels of confirmation bias ($q \ll 1$) may prove to be the most damaging in this regard, as they cause societies to live on a knife-edge where small fluctuations in the information set initially available to the biased agent population can completely censor information signals from opposing viewpoints (see Fig. \ref{fig:curtainplot}). All in all, the model suggests that a \textit{lack} of confirmation bias can ensure that small biased minorities much more easily hijack and dictate public discourse.
	
	The model suggests that as the prevalence of biased agents grows, the available balance of information improves and society is more likely to maintain a long term narrative that is representative of all the information available. On the other hand, it suggests that such societies may grow more polarised. When we examine the net effect of this trade off between bias and polarization through an ensemble approach, our model suggests that the \emph{expected} accuracy of a society may initially improve with the growth of confirmation bias, then reach a maximum at a value $f^*$ before marginal returns to confirmation bias are negative, i.e. confirmation bias experiences an ``optimal'' intermediate value. The model suggests that such value and its corresponding accuracy should increase monotonically with a society's connectivity, meaning that more densely connected societies can support a greater amount of biased reasoners (and healthy debate between biased camps) before partitioning into echo chambers and suffering from polarization.

\section*{SUPPLEMENTARY MATERIALS}	
\section{Section S1: Update dynamics} \label{sec:dynamics}

\subsection{Update dynamics as random matrix $A(t)$.}
		
Consider the set of signals ${s}_i(t)$ possessed by a positively oriented agent $i$ at time $t$ (i.e., $i \in \mathcal{B}^+$). This will consist of a set of signals retained from the previous time step, ${s}_i(t-1)$, and a set of biased signals ${s}^\prime_i(t)$ constructed from the signals available from the nodes $j \in \partial_i$ at the end of time $(t-1)$. 

Let ${s}^*_{i}(t) = \bigcup_{j} {s}_{j}(t-1)$ be the set of the unbiased signals available to node $i$ at time $t$, i.e. the set of nodes $i$ will receive from her neighbors before applying the confirmation bias function. Let $s_i^{*(a)}$ ($a = 1, \ldots, k(k+1)^{t-1}$) be a generic signal in the set ${s}^*_{i}(t)$. After the application of the confirmation bias function, this will be turned into a signal $s^{\prime(a)}_{i}(t) \in {s}^\prime_i(t)$ such that $s^{\prime(a)}_{i}(t) = \pm s^{*(a)}_{i}(t)$ according to the following probabilities:

\begin{eqnarray*} 
\mathrm{Prob}(s'^{(a)}_{i} = +1|s^{*(a)}_{i} = -1) &=& q \\ \nonumber
\mathrm{Prob}(s'^{(a)}_{i} = -1|s^{*(a)}_{i} = -1) &=& (1-q) \\ \nonumber
\mathrm{Prob}(s'^{(a)}_{i} = +1|s^{*(a)}_{i} = +1) &=& 1 \\ \nonumber
\mathrm{Prob}(s'^{(a)}_{i} = -1|s^{*(a)}_{i} = +1) &=& 0 \ . 
\end{eqnarray*}
	
According to the above rules, agent $i$ checks the value of the new incoming signal, and flips it with probability $q$ if it is incongruent with respect to her current orientation. This is entirely equivalent to node $i$ sampling with probability $q$ from the set ${s}^*_{i}(t)$, and with probability $1-q$ from an equally sized set of positive signals belonging to a positively oriented ``ghost'' node. 
	
Let us consider the number $N_i^+(t)$ of positive signals possessed by agent $i$ at time $t$. Due to the above rules, its time evolution will be such that

\begin{equation*}
N_i^+(t) = N_i^+(t-1) + \sum_{j \in \partial_i} \left ( N_j^+(t-1) + w_i(t) N_j^-(t-1) \right ) \ ,
\end{equation*}
where $w_i(t) \in [0,1]$ is a random variable denoting the fraction of negative signals successfully distorted by $i$ of those received by its neighbours at time $t$, with distribution such that $w_i(t) N^-_{\partial_i}(t) \sim \mathrm{Bin}(N^-_{\partial_i}(t),q)$, where $N^-_{\partial_i}(t)$ is the number of negative signals received by $i$ from her neighbourhood at time $t$. When considering agent $i$'s signal mix\footnote{We recall that the signal mix, as per Eq. (1) of the main paper, is defined as the fraction of positive signals possessed by an agent at a certain time, i.e., $x_i(t) = N_i^+(t)/(N_i^+(t) + N_i^-(t)) = N_i^+(t) / (k+1)^t.$}, the above translates to 

\begin{equation} \label{eq:posMR}
x_i(t) = \frac{1}{k+1} \left ( x_i(t-1) + \left (1-w_i(t) \right ) \sum_{j \in \partial_i} x_j(t-1) + w_i(t) k \right ) \ .
\end{equation}

Similarly, for a negatively oriented biased agent (i.e., $i \in \mathcal{B}^-$) we have
\begin{equation} \label{eq:negMR}
x_i(t) = \frac{1}{k+1} \left ( x_i(t-1) + \left (1-w_i(t) \right ) \sum_{j \in \partial_i} x_j(t-1) \right ) \ ,
\end{equation}
with $w_i(t) N^+_{\partial_i}(t) \sim \mathrm{Bin}(N^+_{\partial_i}(t),q)$.

Combining Eqs. \eqref{eq:posMR} and \eqref{eq:negMR} with the time evolution for the signal mix of unbiased agents, which reads 
\begin{equation*} \label{eq:noMR}
x_i(t) = \frac{1}{k+1} \left ( x_i(t-1) + \sum_{j \in \partial_i} x_j(t-1) \right ) \ ,
\end{equation*}
we can see that the time evolution for the vector of signal mixes $x(t)$ can be written as 

\begin{equation} \label{eq:DeGroot}
\boldsymbol{\hat{x}}(t) = \hat{A}(t) \boldsymbol{\hat{x}}(t-1) \ ,
\end{equation}

where $\boldsymbol{\hat{x}} = [\boldsymbol{x}^T, 1, 0]^T$ where the latter terms represent the (fixed) signal mixes of the ghost nodes and the $\boldsymbol{\hat{x}}(t)$ the signal mixes of the original set. $\hat{A}(t)$ is an $(n+2) \times (n+2)$ random matrix with entries with a block structure as follows:

\begin{equation*}
    \hat{A}(t)=
    \left[
    \begin{array}{c|c}
    Q(t) & R(t) \\
    \hline
    0 & I
    \end{array}
    \right] \ ,
\end{equation*}
where $Q(t)$ is an $(n \times n)$ matrix representing the original graph structure with $Q_{ii}(t) = \frac{1}{k+1}$, $Q_{ij}(t) = \frac{1}{k+1}$ where $i$ is a unbiased agent connected to $j$, $Q_{ij}(t) = \frac{1-w_i(t)}{k+1}$ where $i$ is a biased agent connected to $j$, and $0$ otherwise. $R(t)$ is an $(n \times 2)$ matrix representing connections from biased agents to their preferred ghost node (which we index by $+$ and $-$). $R_{i+}(t) = \frac{w_i(t)k}{k+1}$ if $i \in \mathcal{B}^+$ and $0$ otherwise. Analogous weights exist for negatively biased agents to the negative ghost node. $I$ is a $(2 \times 2)$ identity matrix representing the weights of ghost nodes to themselves. $0$ is the $(2 \times n)$ block of zeros representing the (lack of) edges outbound from the ghost nodes.

Finally, it is worth noting that the above formulation consisting of two ghost nodes is fully equivalent to a formulation where each biased agent has a ``personalized'' ghost node that reflects their positive or negative orientation appropriately. In this case $\hat{A}(t)$ is an $(n+fn) \times (n+fn)$ matrix with an extra $fn$ ghost nodes added, one for each biased agent. However, while this formulation has a more favourable interpretation in terms of ``content personalization'', it is less convenient analytically, so for the reminder of the Supplementary Information the simplified ghost node formulation will be utilised.

\subsection{Almost sure convergence of $\hat{A}(t)$.}
\label{sec:conv}

We now proceed to show that stochastic weights $w_i(t)$ appearing in the matrix $\hat{A}(t)$ of \eqref{eq:DeGroot} converge almost surely to $q$ when $t \rightarrow \infty$ as long as at least one signal of each type is held by at least one node in the network. As such the random matrix $\hat{A}(t)$ converges almost surely to a fixed matrix $\hat{A} = \mathbb{E}(\hat{A}(t))$.

Let us consider $i \in \mathcal{B}^+$. As established in the previous section, $w_i(t)$ is simply the fraction of negative signals held by node $i$'s neighbours that $i$ successfully flips to positive at time $t$. Let us also recall that $N^-_{\partial_i}(t)$ represents this set of negative signals available from all $j \in \partial_i$, and that each one is independently flipped to positive with probability $q$. If we can establish that $N^{-}_{\partial_i}(t)$ grows indefinitely as $t \rightarrow \infty$, the Strong Law of Large Numbers (SLLN) can then be invoked to establish the desired result. Since $N^-_{\partial_i}(t) = \sum_{j \in \partial_i} N_j^{-}(t)$, then if $i$'s neighbours possess an increasing and unbounded number of negative signals over time, then $N^-_{\partial_i}(t)$ will also be increasing and unbounded. As such, each $w_i(t)$ will converge almost surely to $q$.

Consider an arbitrary $j \in \partial_i$. Note that since information sets are retained by agents at every time step, we can immediately rule out the possibility of that the number of negative signals held by agent $j$ shrinks over time, and we merely need to show that her set of negative signals does not remain constant over time.	
	
Let us assume that at least one negative signal has been injected into the network at $t=0$, and that one agent $\ell$ possesses such negative signal. In a strongly connected network (such as the $k$-regular network we consider in the main paper), there exists at least one directed path from $k$ to $j$ of length $d$. Let us indicate the probability of a negative signal successfully being transmitted from an agent $a$ to an agent $b$ along such path as $p_{ab}$. We note that $p_{ab} = 1-q$ if $b \in \mathcal{B}^+$ and $p_{ab} = 1$ otherwise. Therefore, the probability of the signal successfully reaching $j$ in $d$ time steps is:

\begin{equation*} 
p_{\ell j} = \prod_{(a,b)} p_{ab} \geq (1-q)^d > 0 \ ,
\end{equation*}

Which allows us to conclude that at each time step $t > d$ there exists a strictly positive probability that a negative signal is added to $j$'s information set. This, in turn, implies that the set of negative signals obtained by $j$ will grow without bound for $t \rightarrow \infty$, which establishes our result. Since this occurs for each $w_i$, we can conclude also that $\hat{A}(t) \xrightarrow{a.s.} \hat{A}$, as well as the block submatrices $Q(t) \xrightarrow{a.s.} Q$ and $R(t) \xrightarrow{a.s.} R$. The edges of these fixed matrices are identical to the structure outlined in the previous section except $w_i$ is replaced with $q$.

Finally it is worth noting that this convergence result depends only on the strong connectedness of $\mathcal{G}$ and not on the edges from the biased agents to the ghost nodes. This is important as this means that even as the orientations of the biased agents change (which is reflected in the rewiring of these ghost node edges), the almost sure convergence is not interrupted.

\section{Section S2: Biased agents settle in their orientation}

In this section, we show that biased agents cannot continue to switch orientation indefinitely, and instead settle into a fixed set of orientations given sufficient time. Recall that a biased agent $i \in \mathcal{B}$ switches her orientation $y_i(t)$ when her information sets switches from a majority of positive signals ($x_i(t) > 1/2$) to a majority of negative signals ($x_i(t) < 1/2$), or vice versa.
			
	We begin by arguing that in some network topologies there exists some $t$ after which biased agents cease switching their orientation. For convenience, we define a network as \textit{settled} at $t^*$ if for all $t > t^*$, $y_i(t^*) = y_i(t)$ for all $i \in \mathcal{B}$. To do this, we first consider an ``adversarial'' toy example designed to maximise the likelihood of indefinite switching, and show that assuming perpetual switching leads to a contradiction even in this case. We then go on to show how other, more complex, network topologies are also guaranteed to settle. We limit to two topologies for brevity but these results can be extended. Alongside the extensive evidence from numerical simulations, we argue that the model is likely to settle for any arbitrary graph.
	
\subsection{Two node network.}

\begin{figure}[h!]
\centering
	\includegraphics[scale=0.2]{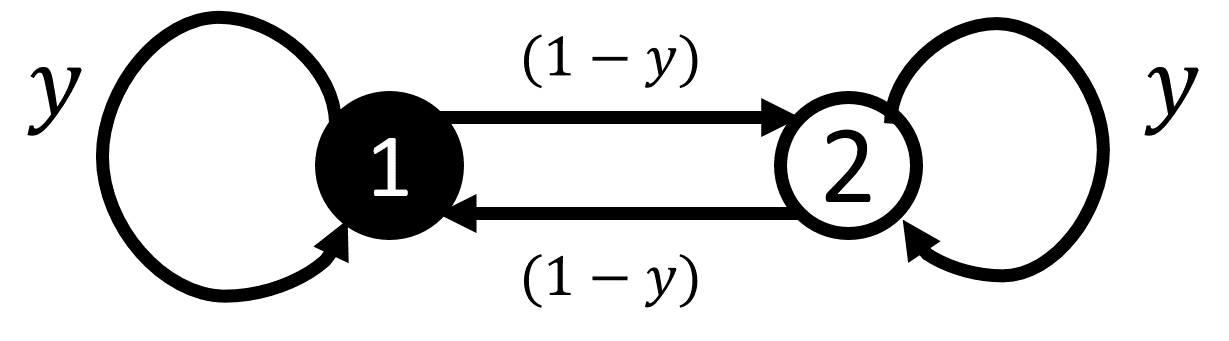}
	\hspace{0.01\textwidth}
	\includegraphics[scale=0.2]{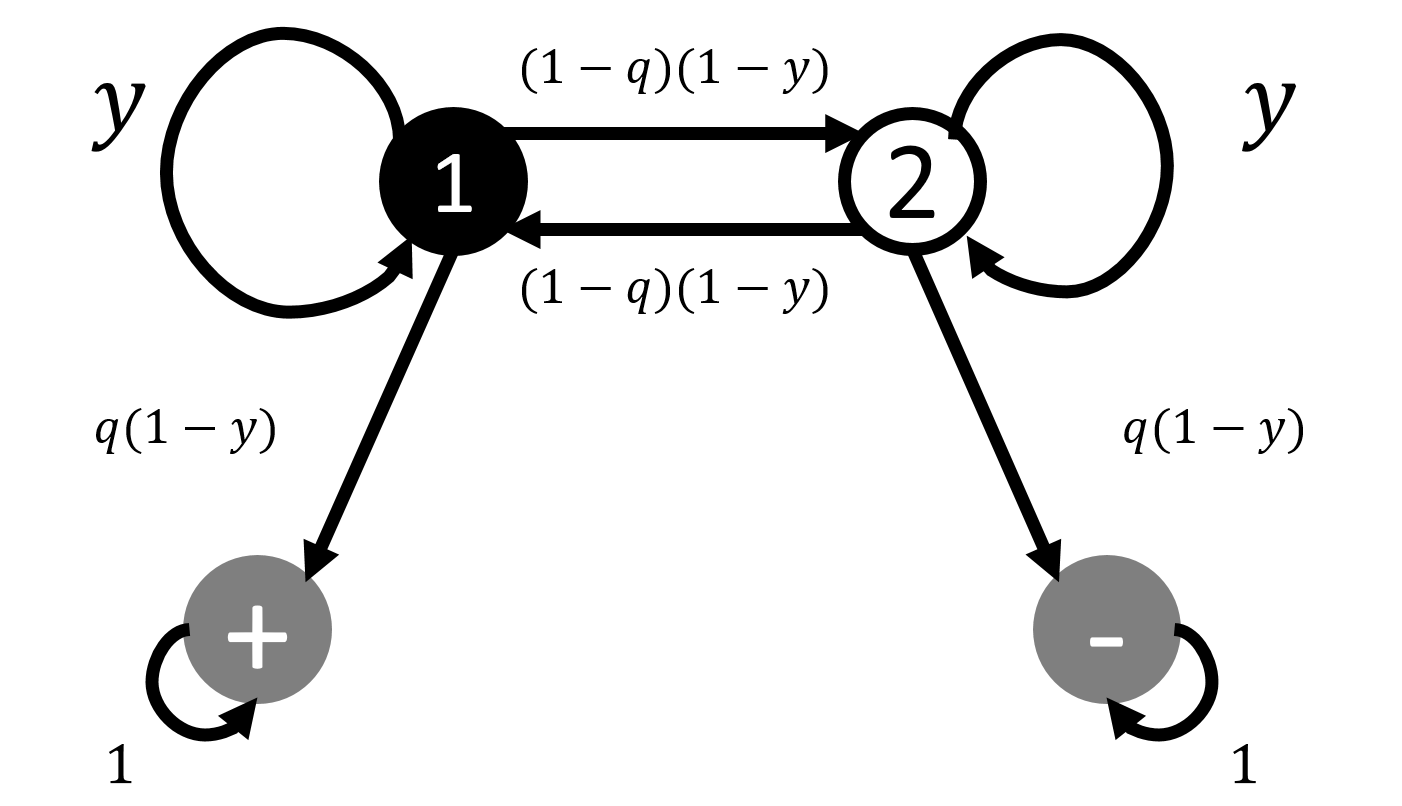}
	\caption{Left: A schematic of the two node symmetric network. Right: A schematic of the two node symmetric network where ghost nodes are introduced to mimic the effect of the biased signals.}
\label{fig:MR1}
\end{figure}

Consider a network with two nodes, labeled $1$ and $2$ respectively, both of which are biased agents. Each node has a self-weight of $y$ and a weight of $(1-y)$ on its sole neighbour\footnote{This setting generalizes the one introduced in Eqs. \eqref{eq:posMR} and \eqref{eq:negMR}, which is recovered for $k=1$ and $y = 1/2$.}. This schematic is illustrated in Figure \ref{fig:MR1}. As has been established in \ref{sec:dynamics}, the signal distortion dyanmics can be mimicked by introducing two ghost nodes that represent a source of positive and negative signals respectively. The weights associated with these ghost nodes are random variables that converge almost surely to $q$ as $t \rightarrow \infty$.
		
	In what follows, we show that this simplified model settles (i.e., both biased agents settle at a finite time on a pair of orientations that they do not thereafter change). For the purposes of illustration, for the moment let us consider the asymptotic case where the random weights have converged to a deterministic set of weights ($q$). 
	
	The outline of this proof (and subsequent ones on alternative network structures) is to establish that in order for a biased agent $i$ to switch orientation, their neighbours must have signal mixes sufficiently far from $i$'s that they can cause $i$ to switch orientation despite the fact that $i$'s ghost node biases her learning to maintain ``inertia'' in the current orientation. However, at the same time, the network structure ensures that nodes tend to converge closely to their neighbourhood, which eventually prevents switching from occurring.
	
\begin{figure}[h!]
\centering
\includegraphics[width=.45\linewidth]{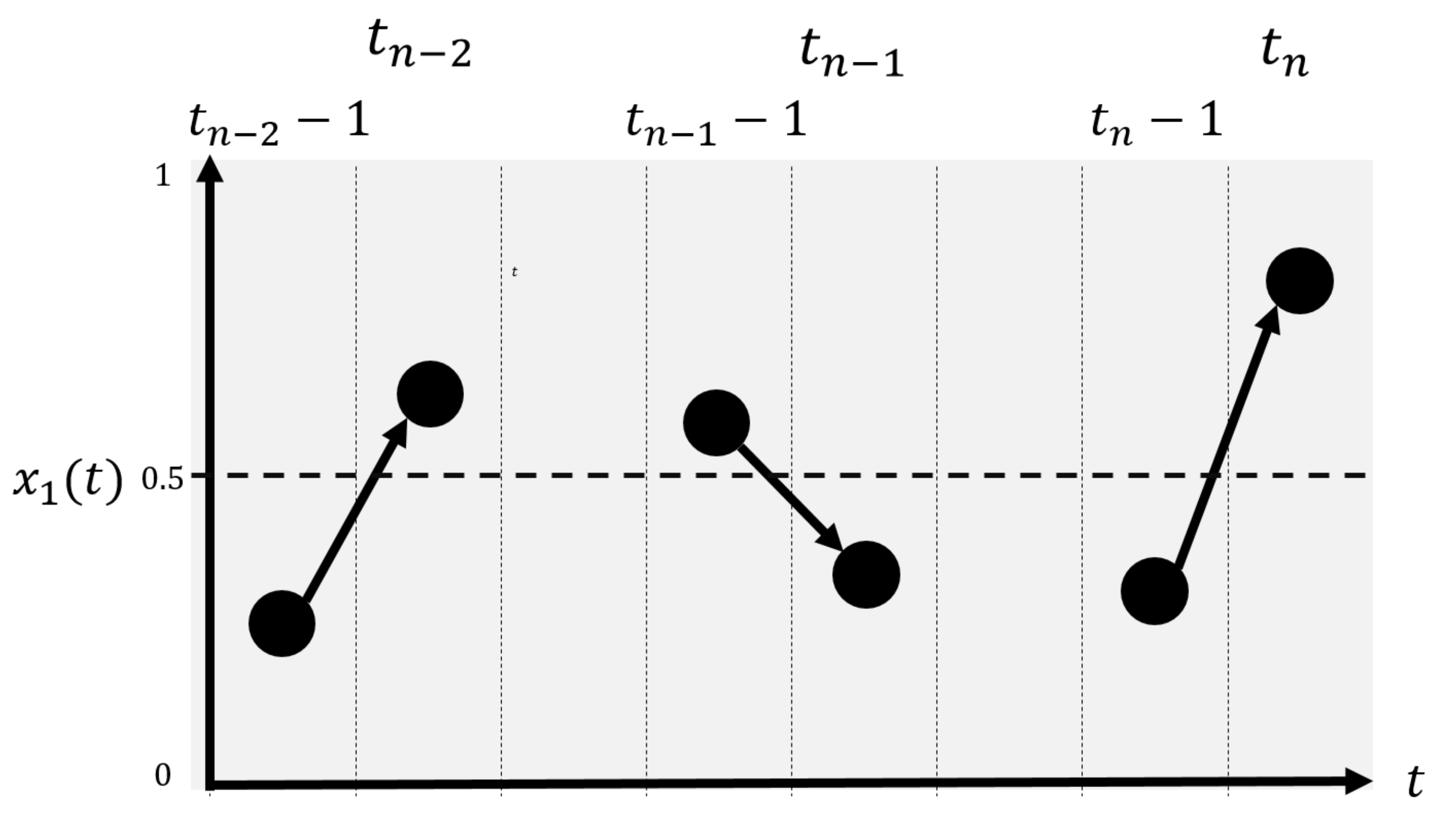}
\includegraphics[width=.45\linewidth]{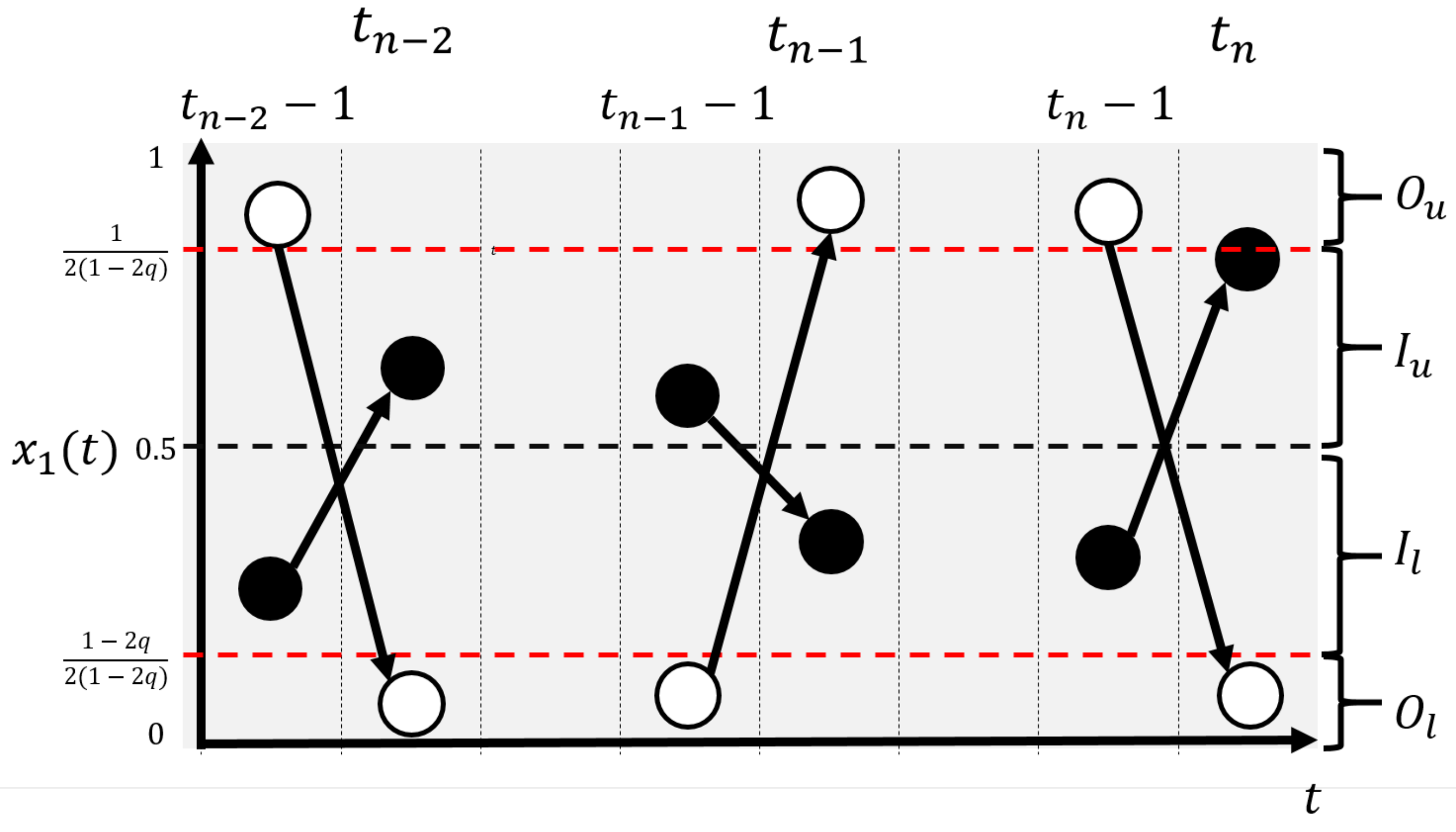}
\caption{By assumption, node $1$ continues to switch orientation at arbitrary time steps $t_{n-2}, t_{n-1}, t_{n}$ by crossing the threshold signal mix $x_i = \frac{1}{2}$. The threshold is denoted by a dashed line. If node $1$ switches, its neighbour (node $2$, white) must cross sufficiently distant thresholds, denoted by the red dashed lines. Furthermore, the switches must be simultaneous, or else the switching terminates perpetually. The regions $\mathcal{O}_u, \mathcal{I}_u, \mathcal{I}_l, \mathcal{O}_l$ are outlined.}
\label{fig:mrswitch}
\end{figure}

	The proof follows by contradiction. Suppose that the model never stabilizes, i.e., that at least one of the biased agents keeps switching perpetually. Suppose node $1$ switches at arbitrary times $\{T\} = \ldots < t_{n-2} < t_{n-1} < t_{n} < \ldots$. We do not assume for now that times in $T$ are over consecutive time steps, the gap between them can be as large as intended (see Fig. \ref{fig:mrswitch}).
	
	Consider some arbitrary $t_n$, where $x_1$ switches from $x_1(t_n - 1)<1/2$ to $x_1(t_n) > 1/2$. Using the model's update rule (see \eqref{eq:negMR}) we can note:
\begin{equation} \label{eq:start1}
x_1(t_n) = yx_1(t_{n}-1) + (1-q)(1-y)x_2(t_{n}-1) > \frac{1}{2} \ ,
\end{equation}
and using the fact that $x_1(t_{n}-1)< 1/2$ we get \ ,
\begin{equation*}
\frac{y}{2} + (1-q)(1-y)x_2(t_{n}-1) \geq \frac{1}{2}
\end{equation*}
which in turn implies
\begin{equation*}
x_2(t_{n}-1) \geq \frac{1}{2(1-q)} > \frac{1}{2} \ .
\end{equation*}

By following the same reasoning one can show that an $x_1$ switch in the opposite direction would imply
\begin{equation*}  \label{eq:bounds2}
x_2(t_{n}-1) \leq \frac{1-2q}{2(1-q)} = \frac{1}{2}-\frac{q}{2(1-q)} < \frac{1}{2} \ .
\end{equation*}

Therefore, for node $1$ to switch endlessly, then node $2$ must also do so, and cannot start from an arbitrary point, but rather has to either be above $1 / (2(1-q))$ or below $(1-2q)/(2(1-q)$ at time $t_n-1$ for $x_1$ to cross the $1/2$ line at time $t_n$ from below or above, respectively. For the sake of convenience we introduce the following regions
\begin{eqnarray*}
\mathcal{O}_l &=& \left[0,\frac{1-2q}{2(1-q)} \right) \\ \nonumber
\mathcal{I}_l &=& \left[\frac{1-2q}{2(1-q)}, \frac{1}{2} \right) \\ \nonumber
\mathcal{I}_u &=& \left[\frac{1}{2},\frac{1}{2(1-q)} \right) \\ \nonumber
\mathcal{O}_u &=& \left(\frac{1}{2(1-q)},1 \right] \ ,
\end{eqnarray*}

Where the subscripts indicate whether the interval lies in the upper or lower hemisphere (above and below $1/2$, denoted by the subscripts $u$ and $l$). We also denote the ``inner'' region $\mathcal{I} = \mathcal{I}_l \cup \mathcal{I}_u$ and the ``outer'' region $\mathcal{O} = [0,1] / \mathcal{I}$ defined by the above boundaries.

According to the above considerations, for node $1$ to switch orientation to negative at $t_n$, then $x_2(t_n-1) \in \mathcal{O}_l$, and for node $1$ to switch to positive at $t_n$, then $x_2(t_n-1) \in \mathcal{O}_u$. These regions are highlighted in \ref{fig:mrswitch}. Clearly, the size of $\mathcal{I}$ grows with $q$ (and $\mathcal{O}$ shrinks with $q$). It is worth noting that for $q>1/2$, the inner region's boundaries exceed $[0,1]$, i.e., orientation switches are impossible. The intuition behind this is that if a node is able to flip more than half of the incongruent signals coming its way, it will never include enough incogruent signals in her information set to switch orientation.
	
We further note that if a node and its neighbour are ever in the same orientation, then any future switches are impossible. Indeed, if two nodes share the same orientation, they are both linked to the same ghost node. As such, the set of available signals for each node is only its neighbour and its ghost node. Regardless of the value of $q$, there is no way for either node to accumulate sufficient incongruent signals to switch orientation. All in all, it follows that both node $1$ and node $2$ must switch at the same time step whenever a switch occurs. This result is illustrated in \ref{fig:mrswitch}.
		
We now show that if $x_2$ lies in the outer region $\mathcal{O}$, it will converge to the inner region $\mathcal{I}$. Furthermore, once it enters the inner region, it cannot leave it. Also, this ceases the switching of the node $1$, since its switching requires $x_2$ to alternate between the upper and lower hemispheres of the outer region.
	
As proved above, at any given time step node $1$ and its neighbour $2$ can either both switch orientation, or both maintain their current orientation. We will consider both possibilities. Assume the former first, in which case we can show the two nodes must grow closer together. Suppose that at time $t_n$, $x_1(t_n)>1/2$, and $x_2(t_n)<1/2$. At $t_n+1$, this orientation switches so $x_2(t_n+1)>x_1(t_n+1)$. Making use of Eqs. \eqref{eq:posMR} and \eqref{eq:negMR}, we can write

\begin{eqnarray} \label{eq:flipshrink}
x_2(t_n+1) - x_1(t_n+1) &=& \left [(1-y)(1-q)-y \right] (x_1(t_n) - x_2(t_n)) - (1-y)q \\ \nonumber
&<& \delta (x_1(t) - x_2(t)) \ ,
\end{eqnarray}
where $\delta = \left [(1-y)(1-q)-y \right] < 1$. Therefore, when the node switches orientation with their neighbour, they must converge strictly closer.\footnote{We require $\delta < 1$ and not $|\delta|<1$. While $\delta<-1$ would violate the convergence criterion, it would also imply $x_2(t+1)<x_1(t+1)$, leading to a contradiction.}
	
We now consider the logical disjunct. Suppose instead that a switch does not occur, and at times $t_n$, $t_n+1$ we have $x_1(t_n),x_1(t_n+1)>1/2$, and $x_2(t_n),x_2(t_n+1)<1/2$. Therefore, we can write

\begin{equation*} \label{eq:convergesame}
x_1(t_n+1) - x_2(t_n+1) = \left [y-(1-y)(1-q)) \right ] (x_1(t_n) - x_2(t_n)) + (1-y)q.
\end{equation*}
If the two nodes are to move closer in this time step, then we must have $x_1(t_n+1) - x_2(t_n+1) < (1-\mu)(x_1(t_n) - x_2(t_n))$ for some $\mu \in (0,1)$. Using this in \eqref{eq:convergesame} we obtain the following sufficient condition for convergence:
\begin{equation} \label{eq:condition1}
x_1(t_n) - x_2(t_n) > \frac{q}{2-q-\frac{\mu}{1-y}} \ .
\end{equation}
Finally, note that if $x_1(t_n) > 1/2$ and $x_2(t_n) \in \mathcal{O}_l$, then:
\begin{equation} \label{eq:condition1_aux}
x_1(t_n) - x_2(t_n) > \frac{1}{2} - \frac{1-2q}{2(1-q)} > \frac{q}{2-q-\frac{\mu}{1-y}}
\end{equation} 
for an arbitrarily small $\mu$. Thus, if $x_2(t_n) \in \mathcal{O}_l$, then the two nodes are sufficiently far apart that the condition in \eqref{eq:condition1} holds, and the two nodes must converge closer together. The parallel argument can be made for the opposite starting orientations.
	
Even if a switch does not occur, then the nodes will converge strictly closer. Indeed, We have established that if $x_2(t) \in \mathcal{O}$, then at each time step the distance $|x_1(t)-x_2(t)|$ must strictly shrink. As such, the nodes will eventually become close enough that $x_2(t) \in \mathcal{I}$, and switching of node $1$ ceases.
	
We complete the proof by showing that once node $2$'s signal mix has entered the inner region $\mathcal{I}$, it cannot leave it. We have established already that nodes must have opposing orientations at all times. Let us consider the case where $x_2(t_n) \in \mathcal{I}_u$ and $x_1(t_n)< 1/2$. Suppose by contradiction that in time step $t_n+1$ node $2$ is able to ``escape'' $\mathcal{I}$ from below, going from below $1/(2(1-q))$ to above such value (i.e., to $\mathcal{O}_u$). This implies
	
\begin{eqnarray*}
\frac{1}{2(1-q)} &<& x_2(t_n+1) = y x_2(t_n) + (1-y)(1-q)x_1(t_n) + (1-y)q \\ \nonumber 
&<& \frac{y}{2(1-q)} + \frac{1}{2}(1-y)(1-q) + (1-y)q \ ,
\end{eqnarray*}
which leads to $1+q < (1-q)^{-1}$, i.e. to the impossible result $q^2 < 0$. Therefore, node $2$ cannot go from $\mathcal{I}$ to $\mathcal{O}_u$. Finally we also know that it cannot go from $\mathcal{I}$ to $\mathcal{O}_l$ as this would require both nodes to switch orientation, which is ruled out because $x_2(t) \in \mathcal{I}$. A parallel argument can be made if the orientations are reversed. Thus, the two node symmetric network will always converge to a region of the signal mix space where the nodes' signal mixes are too close to support any switch of orientation, arriving at the desired result.

The above proof can be easily replicated after relaxing the simplifying asymptotic assumption that $q$ is fixed. This can be done by reintroducing the time-dependent random weights $w_i(t)$ ($i = 1,2$), and recalling that, due to their almost sure convergence to $q$, for any $\epsilon > 0$ there exists a time $t^*$ such that for all $t > t^*$ and for all $i$

\begin{equation*}
q-\epsilon < w_i(t) < q+\epsilon \ .
\end{equation*}
Adjusting the bounds used in the convergence proof to include the above time evolution allows to obtain the same result.
	
\subsection{Star network.}

Let us now consider a $k$-star network of biased agents, with the central node labeled as $0$ and branch nodes labeled as $1, \ldots, k$\footnote{Strictly speaking, the signal diffusion mechanism would need to be modified for non-regular graphs to allow for signal diffusion to be equivalent to node averaging. More complex regular structures can also be shown to converge, but a star graph permits us to show how convergence holds even with a strikingly different topology. We proceed with the star graph for the purpose of illustration.}. As before, allow $y$ to be the self-weight of each node and $q$ the confirmation bias parameter. Assume for simplicity that the central node has a weight of $(1-y) / k$ on each branch node.
	
Firstly, note that if any branch node switches indefinitely, then the central node $0$ must also switch indefinitely (or else there would be no ``driving force'' causing the branch nodes to switch). So, let us focus on showing that it is impossible for the central node to do so. The logic of the two-node network proof can be followed almost exactly by replacing $x_1(t)$ and $x_2(t)$  with $x_0(t)$ and $\sum_{j=1}^k x_j(t) / k$, respectively.

The first set of results up to \eqref{eq:bounds2} follow precisely given the substitution of terms above. We use this to establish once again that for $x_0(t)$ to switch indefinitely $\sum_{j=1}^k x_j(t)/k$ must oscillate between $1/(2(1-q))$ and $(1-2q)/(2(1-q))$, i.e., between the upper and lower hemispheres of the outer region $\mathcal{O}$. Furthermore, whenever the central node switches orientation, the branch nodes' average signal mix must also change from above to below $1/2$ (or vice versa), even if none of the branch nodes in particular switch orientation.
	
The next steps follow closely those of the two-nodes network. Suppose firstly that the central node switches orientation (and the branch nodes' average must also shift accordingly). Suppose that at time $t_n$, $x_0(t_n)> 1/2$, and $\sum_{j=1}^k x_j(t_n)/k<1/2$. At $t_n+1$, this orientation switches so that $\sum_{j=1}^k x_j(t_n+1)/k >x_0(t_n+1)$. Adapting Eqs. \eqref{eq:posMR} and \eqref{eq:negMR} to the present case, we have
	
\begin{eqnarray*}
\frac{1}{k} \sum_{j=1}^k x_j(t_n+1) - x_0(t_n+1) &=& \left [\frac{1}{k}\sum_{j=1}^k \left (x_j(t_n) + (1-y)(1-q) x_0(t_n) + (1-y)q g_j(t_n) \right) \right] \\ \nonumber
&-& \left [y x_0(t_n) + (1-y)(1-q) \frac{1}{k} \sum_{j=1}^k x_j(t_n) + (1-y)q \right ] \ ,
\end{eqnarray*}
where we have introduced a new indicator variable such that $g_j(t) = 1$ is node $j$ is positively oriented at time $t$, and $g_j(t) = 0$ otherwise. Let $g(t) = \sum_{j=1}^k g_j(t) / k$ be the fraction of positively oriented branch nodes. We can then simplify the above:

\begin{eqnarray*}
\frac{1}{k} \sum_{j=1}^k x_j(t_n+1) - x_0(t_n+1) &=& \left [ (1-y)(1-q)-y \right ] \left (x_0(t_n) - \frac{1}{k} \sum_{j=1}^k x_j(t_n) \right) - (1-y)q(1-g(t_n)) \\ \nonumber
&<& \left [ (1-y)(1-q)-y \right ] \left (x_0(t_n) - \frac{1}{k} \sum_{j=1}^k x_j(t_n) \right) \ ,
\end{eqnarray*}
where we used the fact that $(1-y)q(1-g(t_n)) > 0$. This can be rewritten as 

\begin{equation*}
\frac{1}{k} \sum_{j=1}^k x_j(t_n+1) - x_0(t+1) < \delta \left (x_0(t_n) - \frac{1}{k} \sum_{j=1}^k x_j(t_n) \right ) \ ,
\end{equation*}
where $\delta = \left [ (1-y)(1-q)-y \right ] < 1$, which re-establishes the result of \eqref{eq:flipshrink}: if the central node flips, it must converge strictly closer to the branch nodes.
\begin{figure}[h!]
\centering
\includegraphics[width=0.4\textwidth]{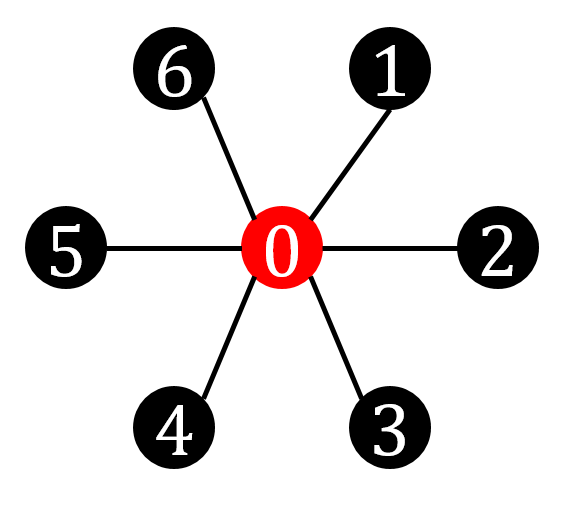}
\caption{A network structure consisting of a central node $0$ and $k = 6$ branch nodes. All nodes are biased agents for the purposes of the toy example.}
\label{fig:star}
\end{figure}
Next, we establish that in the time steps where the central node does not switch orientation, the centre and branches still converge as long as the branch average is within $\mathcal{O}$. The reasoning follows the one of the previous section exactly given the appropriate substitutions, and we can replace the condition in \eqref{eq:condition1} with:
\begin{equation} \label{eq:condition2}
x_0(t_n) - \frac{1}{k} \sum_{j=1}^k x_j(t) > \frac{q(1-g(t_n))}{2-q-\frac{\mu}{1-y}} \ .
\end{equation}
Recall that $x_0(t_n) > 1 / 2$ and $\sum_{j=1}^k x_j(t) / k  < (1-2q)(2(1-q)) \in \mathcal{O}_l$, therefore:
\begin{equation*}
x_0(t_n) - \frac{1}{k} \sum_{j=1}^k x_j(t) > \frac{1}{2} - \frac{1-2q}{2(1-q)} > \frac{q}{2-q-\frac{\mu}{1-y}} \geq \frac{q(1-g(t))}{2-q-\frac{\mu}{1-y}}
\end{equation*}
for an arbitrarily small $\mu > 0$. Hence, even if a switch does not occur, then the nodes will converge strictly closer.
	
The final steps of the proof mirror those that follow \eqref{eq:condition1_aux} of the previous section, except a factor of $g(t_n)$ dampens the ability of the branch nodes to escape the inner region even further. As such, we establish that even on a star network structure, the biased agents cannot switch their orientation endlessly, and must eventually converge.

\subsection{Simulated dynamics and convergence criteria.}
	
As has been established in the previous sections, settling is guaranteed under some simple network topologies chosen specifically to hinder convergence. We round out the argument by noting that settling also occurs in simulations for the $k$-regular network employed throughout the paper and in the following proofs.
	
In what follows, we establish criteria for the case of a fixed $q$. Analogous criteria can be easily established for the case of stochastic convergent weights $w_i(t)$ instead, although without much adding much insight. Furthermore, in practice the stochasticity rapidly settles in numerical simulations, meaning that convergence can be safely studied using the asymptotic fixed $q$ assumption.
	
In order to guarantee that a network has in fact settled over the course of a simulation, we identify a ``settling'' rule for the signal mix ${\boldsymbol{\hat{x}}}(t)$. As we demonstrate in the following section, if one assumes that the biased agents at time $t$ no longer switch orientations, one can calculate the steady state that would arise from this configuration of biased agents. Call this $\boldsymbol{\hat{x}}^*({\boldsymbol{\hat{x}}}(t))$. We can show that if the signal mix ${\boldsymbol{\hat{x}}}(t)$ is sufficiently close to its corresponding steady state $\boldsymbol{\hat{x}}^*({\boldsymbol{\hat{x}}}(t))$ it will converge uniformly to that steady state without any further changes to any agent's orientation.

Define the difference between a signal mix and its steady state:

\begin{equation*} 
\epsilon(t) = \boldsymbol{\hat{x}}(t) - \boldsymbol{\hat{x}}^* \ . 
\end{equation*}
Recalling that the model's dynamics is such that 	
	\begin{equation*} 
	{\boldsymbol{\hat{x}}}(t) = \hat{A}{\boldsymbol{\hat{x}}}(t-1) \
	\end{equation*}
we then have	
	\begin{equation*} 
	{\boldsymbol{\hat{x}}}^* + \epsilon(t) = \hat{A}{\boldsymbol{\hat{x}}}^* + \hat{A}\epsilon(t-1) = {\boldsymbol{\hat{x}}}^* + \hat{A}\epsilon(t-1) \ ,
	\end{equation*}
and therefore:	
	\begin{equation*} 
	\epsilon(t) = \hat{A}\epsilon(t-1) \ .
	\end{equation*}
Finally, define $\epsilon^*(t) = \max_i(|\epsilon_i(t)|) = ||\epsilon(t)||_{\infty}$. Then for any arbitrary biased node:
	
	\begin{equation*} 
	\epsilon_i(t+1) = \hat{a}_{ii}\epsilon_i(t) + (1-q)\sum_j \hat{a}_{ij}\epsilon_j(t) \ ,
	\end{equation*}
where $\hat{a}_{ij}$ is the weight between node $i$ and $j$ in matrix $\hat{A}$, and we use the fact that the ghost nodes are always at their exact steady state, so their $\epsilon_G = 0$. Then taking the absolute distance and using the triangular inequality:
	
\begin{eqnarray*} 
&& |\epsilon_i(t+1)| \leq \hat{a}_{ii}|\epsilon_i(t)| + (1-q)\sum_j \hat{a}_{ij}|\epsilon_j(t)| \\ \nonumber
&& \leq \hat{a}_{ii}|\epsilon^*(t)| + (1-q)\sum_j \hat{a}_{ij}|\epsilon^*(t)| = (1 - q(1 - \hat{a}_{ii}))|\epsilon^*(t)| < |\epsilon^*(t)| \ ,
\end{eqnarray*}
and similarly for an unbiased agent, we can show:
\begin{equation*} 
|\epsilon_i(t+1)| \leq |\epsilon^*(t)| \ .
\end{equation*}
	
In short, for each steady state once the current signal mixes are within some $\epsilon$-cube of the steady state, they must remain within that $\epsilon$-cube. Furthermore, biased agents at each time step must converge strictly closer to the steady state. A larger $q$ or smaller self-weight ($\hat{a}_{ii}$) will cause faster convergence.
	
	Finally, we can also note because the network is strongly connected, there are some $r \geq 1$ steps between each unbiased and a biased node, and so it can be shown that in a finite number of steps \textit{all} nodes must converge strictly closer to the steady state than the maximum threshold of the $\epsilon$-cube.
	
	Given all the above, we can now explicitly outline a ``stable'' region. Denote:
	
	\begin{equation*} 
	\epsilon_s = \min_i \left (|x^*_i - \frac{1}{2}| \right) \ .
	\end{equation*}
That is, the closest any of the steady states are to the threshold. If the current signal mixes can get within the $\epsilon_s$-ball of the steady state, then there can be no crossing the $0.5$ threshold and as such the steady state cannot move - this is a sufficient condition for settling to be guaranteed.
	
	Using this condition, we are able to demonstrate settling occurs for a wide range of parameters for $k$-regular networks. We tested the condition on $1000$ iterations each of the following parameter sets: $n=10^3$, $p=0.5$, $k =\{3, 4, 5, 10, 100, 999\}$, $f = \{.05, .1, .2, .5, 1\}$, $q=\{.05, .1, .25, .5, .75. 1\}$. The settling criteria was successfully reached for every single run of the model, establishing extremely high confidence that the $k$-regular biased information aggregation model always settles.

\section{Section S3: Convergence of signal mixes}
	
	So far, we have established that the random update matrix $A(t)$ converges almost surely to a fixed update matrix $A$. Furthermore, we have demonstrated with extremely high confidence that biased agents settle in their orientation after some finite time. As such, for biased $k$-regular networks, assume that there exists some time $t^*$ after which biased agents cease switching their orientation. Define $\hat{y}_{\mathcal{B}}^*$ as the steady state fraction of positively oriented biased agents. Then the following holds.
	
	\begin{itemize}
	
    \item[(1)] The signal mix vector $\boldsymbol{\hat{x}}(t)$ converges to some $\boldsymbol{\hat{x}}^* = \hat{A}^*\boldsymbol{\hat{x}}(0)$ for both biased and unbiased networks, where $\hat{A}^*$ is a steady-state matrix of influence weights which can be computed explicitly.
	
    \item[(2)] Unbiased networks achieve consensus, and converge to influence weights of $a^*_{ij} = 1/n$ for all pairs $(i,j)$. This ensures that, for all $i \in V$, $x^*_i=x^*_V=\bar{x}(0)$, where $\bar{x}(0) = \sum_{i=1}^n s_i$ is the intial average signal mix.
	
    \item[(3)] Biased networks where $\hat{y}_{\mathcal{B}}^* = 0,1$ achieve consensus, and converge to influence weights $\hat{a}^*_{ij} = 0$ for all pairs $(i, j) \in V$, $\hat{a}^*_{i+} = \hat{y}_{\mathcal{B}}^*$ and $\hat{a}^*_{i-} = 1-\hat{y}_{\mathcal{B}}^*$ for all $i \in V$.
	
    \item[(4)] Biased networks where $ 0 < \hat{y}_{\mathcal{B}}^* < 1$ do not achieve consensus, and converge to influence weights $\hat{a}^*_{ij} = 0$ for all $(i, j) \in V$, and $\hat{a}^*_{i+} + \hat{a}^*_{i-}=1$ for all $i \in V$.
    \end{itemize}    
	
	We note that results regarding unbiased networks (part of (1) and all of (2)) are already well established results (see, for example, \cite{degroot1974reaching}) and are listed purely for comparison with biased networks. We focus on proving the remainder of the results.
	
	The results follow from the structure of $\boldsymbol{\hat{x}}^* = \hat{A}^*\boldsymbol{\hat{x}}(0) = \lim\limits_{t \to \infty}\prod_{\tau=0}^{t}\hat{A}(\tau){\boldsymbol{\hat{x}}}(0)$. We proceed by demonstrating that despite the stochasticity in the random update mechanism $\hat{A}(\tau)$, the steady state converges to a fixed vector $\boldsymbol{\hat{x}}^*$.
	
	First note that for $\tau > t^*$ the biased agents will have ceased switching their orientation, and the random update matrix $\hat{A}(\tau)$ will have a fixed underlying structure $\hat{A} = \mathbb{E}(\hat{A}(\tau))$. The proof will follow by demonstrating that $\lim\limits_{t \to \infty}\prod_{\tau=0}^{t}\hat{A}(\tau) = \lim\limits_{t \to \infty}\hat{A}^t$. That is, the products of random matrices converges to the products of their expectation.
	
	Firstly recall the block structure of $\hat{A}(\tau)$:

\begin{equation*}
    \hat{A}(\tau)=
    \left[
    \begin{array}{c|c}
    Q(\tau) & R(\tau) \\
    \hline
    0 & I
    \end{array}
    \right] \ ,
\end{equation*} 
with dimensions (clockwise from top-left): $(n \times n)$, $(n \times 2)$, $(2 \times 2)$, $(2 \times 2)$. Important properties of the blocks include:

\begin{eqnarray*}
    && Q(\tau) = Q + \epsilon_Q(\tau) \xrightarrow{a.s.} Q \\ \nonumber
    && R(\tau) = R + \epsilon_R(\tau) \xrightarrow{a.s.} R \ .
\end{eqnarray*}
The properties above indicate that the blocks converge to their deterministic counterparts almost surely. This allows us to state that for any $\epsilon$ and matrix norm $||.||$, there is guaranteed some $t' \geq t^*$ such that for all $\tau > t'$, $||Q-Q(\tau)|| = ||\epsilon_Q(\tau)|| <\epsilon$. Also, the matrix $Q$ is such that

\begin{eqnarray*}
    && \sum_j Q_{ij} < 1 \ \ \forall i \in \mathcal{B} \\ \nonumber
    && \sum_i Q_{ij} < 1 \ \ \forall j \in \partial\mathcal{B} \ .
\end{eqnarray*}
The properties respectively indicate that the limit matrix $Q$ is both row and column sub-stochastic. Row sub-stochasticity follows from the outgoing edges from the set of biased agents ($\mathcal{B}$). For the $k$-regular graphs that are the focus of our analysis this can be specifically shown to be $\frac{(1-q)k+1}{k+1}$. Column sub-stochasticity follows from the neighbours of the biased agents ($\partial \mathcal{B}$) having incoming connections necessarily less than $1$.

We now define the product of the \textit{random} matrices $\hat{A}(\tau)$ as:

\begin{equation*}
    \prod_{\tau=0}^{t}\hat{A}(\tau) = \tilde{A}(t,0) = \left[
    \begin{array}{c|c}
    \tilde{Q}(t,0) & \tilde{R}(t,0) \\
    \hline
    0 & I
    \end{array}
    \right] \ ,
\end{equation*}
where $\tilde{Q}(t,0$ and $\tilde{R}(t,0)$ are placeholder terms for the the random block matrix products which arise through products of the random matrices $\hat{A}(\tau)$. Consider also the deterministic analog to this expression:

\begin{equation*}
    \prod_{\tau=0}^{t}\hat{A} = \hat{A}^t = \dot{A}(t,0) = \left[
    \begin{array}{c|c}
    \dot{Q}(t,0) & \dot{R}(t,0) \\
    \hline
    0 & I
    \end{array}
    \right] \ .
\end{equation*}
This formulation defines a random and analogous deterministic sequence for each of the blocks, denoted by $\tilde{A}(t,0)$ and $\dot{A}(t,0)$ respectively.

Firstly, we demonstrate that $\lim\limits_{t \to \infty}\dot{Q}(t,0) = \lim\limits_{t \to \infty}Q^t = \textbf{0}$. Consider the $2$-norm $||.||_2$. We consider first the deterministic matrix $Q$. Recall that since $Q$ is doubly sub-stochastic, $Q^T Q$ is necessarily sub-stochastic and therefore:
    
    \begin{flalign*}
        ||Q|| =  (1 - \delta) < 1 \ ,
    \end{flalign*}
for some $\delta \in (0,1)$. It follows that:
    
    \begin{flalign*}
        \lim\limits_{t \to \infty}||Q^t|| \leq \lim\limits_{t \to \infty}||Q||^t  = \lim\limits_{t \to \infty}(1 - \delta)^t = 0 \ ,
    \end{flalign*}
    therefore $\lim\limits_{t \to \infty}Q^t = \textbf{0}$ (making use of the fact that $||X|| = 0 \iff X = \textbf{0}$). Now consider the term of interest $\tilde{Q}(t,0)$:
    
    \begin{flalign*}
        ||\tilde{Q}(t,0)|| = ||\prod_0^t Q(\tau)|| \leq \prod_0^t||Q(\tau)|| \ .
    \end{flalign*}
Note that $||Q(\tau)|| \leq 1$ for all $\tau$. However we can show that almost all $||Q(\tau)|| < 1$:
    
    \begin{flalign*}    
        ||Q(\tau)|| = ||Q + \epsilon_Q(\tau)|| \leq ||Q|| + ||\epsilon_Q(\tau)||
    = (1 - \delta) + ||\epsilon_Q(\tau)||
    \end{flalign*}
    
    We select some $t'$ such that for all $\tau > t',  ||\epsilon_Q(\tau)|| = \mu < \delta$ for some $\mu > 0$. Therefore: 
    
    \begin{equation} \label{eq:qnorm}
        ||Q(\tau)|| \leq (1-\delta+\mu) = (1-\delta^*) < 1 \ ,
    \end{equation}
where $\delta^* = \delta+\mu$. We can now conclude:
    
    \begin{flalign*}
        \lim\limits_{t \to \infty}||\tilde{Q}(t,0)|| \leq \lim\limits_{t \to \infty}\prod_{t^*}^t (1-\delta^*) = \lim\limits_{t \to \infty}(1 - \delta^*)^t = 0 \ .
    \end{flalign*}
    

    We now show that $\lim\limits_{t \to \infty}\tilde{R}(t,0) = \lim\limits_{t \to \infty}\dot{R}(t,0) = (I-Q)^{-1}R$. Consider firstly the deterministic sequence, which can be defined through the following iterative relationship:
    \begin{equation} \label{eq:affine}
        \dot{R}(t,0) = Q \dot{R}(t-1,0) + R \ .
    \end{equation}
Note again that $\dot{R}(t,0)$ refers to the $t$-th term in a deterministic sequence whereas $Q$ and $R$ are specific block matrices. The expression \eqref{eq:affine} can be straightforwardly solved in the limit:
    
    \begin{equation*}
        \lim\limits_{t \to \infty}\dot{R}(t,0) = (I-Q)^{-1}R \ .
    \end{equation*}
Consider now the random sequence, which can be defined analogously:
    
    \begin{equation} \label{eq:randaffine}
        \tilde{R}(t,0) = Q(t) \tilde{R}(t-1,0) + R(t) \ .
    \end{equation}
    Note here $\tilde{R}(t,0)$ refers to the $t$-th term in a random sequence and $Q(t)$ and $R(t)$ are random block matrices that occur at time $\tau = t$. In order to proceed we define:
    
    \begin{equation} \label{eq:refdiff}
        \tilde{R}(t,0) = \dot{R}(t,0) + E(t) \ ,
    \end{equation}
    where here $E(t)$ is an error term capturing the difference between the terms of the deterministic and random sequences at time $\tau = t$. We substitute \eqref{eq:refdiff} into \eqref{eq:randaffine}:
    
    \begin{equation} \label{eq:combaffine}
        \tilde{R}(t,0) = Q(t) (\dot{R}(t-1,0) + E(t-1)) + R(t) \ .
    \end{equation}
    We now substitute the definition of $Q(\tau)$ and $R(\tau)$ into \eqref{eq:combaffine}:
    
    \begin{equation*}
\tilde{R}(t,0) = Q \dot{R}(t-1,0) + R + Q E(t-1) + \epsilon_Q(t)(\dot{R}(t-1,0)+E(t-1)) + \epsilon_R(t) \ .
    \end{equation*}
    Note that we can substitute \eqref{eq:affine} for the two leading terms on the RHS:
    
    \begin{equation*}
\tilde{R}(t,0) - R(t,0) = Q E(t-1) + \epsilon_Q(t)(\dot{R}(t-1,0)+E(t-1)) + \epsilon_R(t) \ .
    \end{equation*}
   We can now take the 2-norm $||.||_2$:
    
    \begin{flalign*}
        ||E(t)|| = ||\tilde{R}(t,0) - \dot{R}(t,0)|| = ||Q E(t-1) + \epsilon_Q(t)(\dot{R}(t-1,0)+E(t-1)) + \epsilon_R(t)|| \\
        \leq ||Q|| \ ||E(t-1)|| + ||\epsilon_Q(t)|| \ ||\tilde{R}(t-1)|| + ||\epsilon_R(t)|| \ .
    \end{flalign*}
We can now substitute in \eqref{eq:qnorm} and once again make use of the fact that for any $0<\epsilon$ we can define $t'$ such that $||\epsilon_Q(t)||$, $||\epsilon_R(t)|| < \epsilon$. We also note that $||\tilde{R}(t-1)|| < n$ (where $n$ is the size of the network), therefore
      \begin{equation*}
        ||E(t)|| \leq (1-\delta)||E(t-1)|| + \epsilon (n+1) \ ,
    \end{equation*}
and therefore for a sufficiently small $\epsilon$, there is a corresponding $t'$ such that for $t > t'$:
    
    \begin{equation*}
        ||E(t)|| \leq (1-\delta^*)||E(t-1)|| < ||E(t-1)|| \ .
    \end{equation*}
    Finally we get:
    
    \begin{equation*}
        \lim\limits_{t \to \infty} ||\tilde{R}(t,0)-\dot{R}(t,0)|| = \lim\limits_{t \to \infty}||E(t)|| = 0 \ ,
    \end{equation*}
    which allows us to conclude that $\lim\limits_{t \to \infty} \tilde{R}(t,0) = \lim\limits_{t \to \infty} \dot{R}(t,0) = (I-Q)^{-1}R$.
    
    
    We can combine these results to conclude:
    
    \begin{equation*}
        \hat{A}^* = \lim\limits_{t \to \infty} \prod_{\tau=0}^{t}\hat{A}(\tau)
        = \lim\limits_{t \to \infty} \tilde{A}(t,0)
        = \left[
        \begin{array}{c|c}
        \lim\limits_{t \to \infty} \tilde{Q}(t,0) & \lim\limits_{t \to \infty} \tilde{R}(t,0) \\
        \hline
        0 & I
        \end{array}
        \right]
        = \left[
        \begin{array}{c|c}
        0 & (I-Q)^{-1}R \\
        \hline
        0 & I
        \end{array}
        \right] \ .
    \end{equation*}
    Given that $\boldsymbol{\hat{x}}^* = \hat{A}^*\boldsymbol{\hat{x}}(0)$ we can conclude Result (1) - that the signal mixes do converge. In particular:
	
	\begin{equation*}
	    \boldsymbol{\hat{x}}^* = \hat{A}^* \boldsymbol{\hat{x}}(0)
        = \left[
        \begin{array}{c|c|c}
        0 & (I-Q)^{-1}R^+ & (I-Q)^{-1}R^- \\
        \hline
        0 & 1 & 0 \\
        \hline
        0 & 0 & 1
        \end{array}
        \right]
         \left[
        \begin{array}{c}
        \boldsymbol{\hat{x}}(0) \\
        \hline
        1 \\
        \hline
        0
        \end{array}
        \right]
        = \left[
        \begin{array}{c}
        (I-Q)^{-1}R^+ \\
        \hline
        1 \\
        \hline
        0
        \end{array}
        \right] \ ,
    \end{equation*}
that is, the steady state signal mixes of the agents not a function of the initial signals $\hat{x}(0)$. The signal mixes are instead entirely a function of the steady state orientations of the biased agents, encoded by the vector $R^+$, the edges from the positive biased agents to the positive ghost nodes.
	
	Our remaining conclusions follow summarily from this. If all biased agents are negative ($\hat{y}_\mathcal{B}^* = 0$) $R^+$ is $0$ and $x^*$ is $0$ for all agents. Inversely if all biased agents are positive ($\hat{y}_\mathcal{B}^* = 1$), $x^*$ is $1$ for all agents. For any other configuration of biased agents, the steady state is determined by the closed form $(I-Q)^{-1}R^+$. In this scenario $z_V^* > 0$ trivially as some biased nodes will be of the minority orientation. However, more crucially $z_R^* \geq 0$. That is, unbiased agents are no longer guaranteed to converge despite having no bias mechanism themselves. We investigate this and other properties of the unbiased agents in more detail in the next section.
	
\section{Section S4: Steady state signal mix distribution}
	
We now seek to approximate the distribution of signal mixes of the agents once the steady state is reached. We will first approximate the average steady state signal mix of each sub-population in the network, followed by the steady state signal mix variance, and finally the full distribution itself. We will do this for the $k$-regular network case used in the body of the paper, and show via numerical simulations that it also captures the model's dynamics on more heterogeneous networks. Let us note that the results given in the following are the empirical distribution of the signal mixes for a given run of the model, as opposed to an ensemble over all possible runs of the model.

\subsection{Steady state expected signal mix.}
	
As detailed in \ref{sec:dynamics}, the model converges to a steady state ${\hat{\textbf{x}}}^*$ which is entirely contingent on the settled orientation of the biased agents in the network. In what follows, we will calculate an approximation for the model's steady state expected signal mix \textit{conditional} on a given fraction of positively oriented biased agents $f^+(t) = \hat{y}_{\mathcal{B}}(t) f$. We will then show how under some reasonable assumptions the ``settled'' value of $f^+(t)$ can be approximated from the initial orientation $f^+(0)$.
	
Consider an agent $i$ picked uniformly at random at time $t$ from the unbiased, positively oriented biased, negatively oriented biased sub-populations. Let us denote the signal mixes of agents belonging to such sub-populations as $\hat{x}_{i_{\mathcal{U}}}(t)$, $\hat{x}_{i_{\mathcal{B}^+}}(t)$ and $\hat{x}_{i_{\mathcal{B}^-}}(t)$, respectively. We are interested in establishing the expected steady state values for each of these quantities, denoted as
\begin{equation*}
\hat{x}_{\mathcal{U}}^* = \lim_{t \rightarrow \infty} \mathbb{E} [\hat{x}_{i_{\mathcal{U}}}(t)] = \lim_{t \rightarrow \infty} \hat{x}_{\mathcal{U}}(t) \ ,
\end{equation*}
with analogous definitions for $\hat{x}_{\mathcal{B}^+}^*$ and $\hat{x}_{\mathcal{B}^-}^*$.

We begin by considering the sub-population of unbiased agents at some finite $t$. We note the following:
\begin{equation} \label{eq:xU}
\hat{x}_{\mathcal{U}} (t+1) = \mathbb{E} [\hat{x}_{i_{\mathcal{U}}}(t+1)] = \mathbb{E} \left [\frac{\hat{x}_{i_{\mathcal{U}}}(t)+\sum_{j \in \partial_i}\hat{x}_j(t)}{k+1} \right ] = 
\frac{\hat{x}_{\mathcal{U}}(t)+k\mathbb{E} [\hat{x}_j(t)]}{k+1} \ ,
\end{equation}
where $\mathbb{E}[\hat{x}_j(t)]$ refers to the expected signal mix of a randomly picked agent $j$ from the entire population, which of course consists of the three aforementioned sub-populations. Therefore, we have
	
\begin{eqnarray} \label{eq:generic_agent}
\mathbb{E} [\hat{x}_j(t)] &=& (1-f) \mathbb{E} [\hat{x}_j(t) | j \in U] + f^+(t) \mathbb{E} [\hat{x}_j(t) | j \in \mathcal{B}^+(t)] + f^-(t) \mathbb{E} [\hat{x}_j(t) | j \in \mathcal{B}^-(t)] \\ \nonumber
&=& (1-f) \hat{x}_{\mathcal{U}}(t) + f^+(t) \hat{x}_{\mathcal{B}^+}(t) + f^-(t) \hat{x}_{\mathcal{B}^-}(t) \ .
\end{eqnarray}

Plugging the above in \eqref{eq:xU} we get
\begin{equation*}
\hat{x}_{\mathcal{U}}(t+1) = \frac{1}{k+1} \left [ (1+k(1-f))\hat{x}_{\mathcal{U}}(t) + kf^+\hat{x}_{\mathcal{B}^+}(t) + kf^-\hat{x}_{\mathcal{B}^-}(t) \right ] \ .
\end{equation*}
	
Repeating the above steps for positively oriented biased agents we get
\begin{eqnarray*}
\hat{x}_{\mathcal{B}^+}(t+1) &=& \mathbb{E} [\hat{x}_{i_{\mathcal{B}^+}}(t+1)] = \mathbb{E} \left [\frac{\hat{x}_{i_{\mathcal{B}^+}}(t)+\sum_{j \in \partial_i} \left ((1-w_i(t))\hat{x}_j(t) + w_i(t) \right)}{k+1} \right ] \\ \nonumber
&=& \frac{\hat{x}_{\mathcal{B}^+}(t)+(1-q)k\mathbb{E}[\hat{x}_j(t)] + kq}{k+1} \ .
\end{eqnarray*}
where we have explicitly referenced the random variable $w_i(t)$ representing the fraction of successfully distorted negative signals (see \ref{sec:dynamics}), and made use of the fact that $\mathbb{E}[w_i(t)] = q$. We can use \eqref{eq:generic_agent} again and write
\begin{equation*}	
\hat{x}_{\mathcal{B}^+}(t+1) = \frac{1}{k+1} \left [ k(1-f)(1-q)\hat{x}_{\mathcal{U}}(t) + ((1-q)kf^+(t) + 1)\hat{x}_{\mathcal{B}^+}(t) + (1-q)kf^-(t)\hat{x}_{\mathcal{B}^-}(t) + kq \right ] \ ,
\end{equation*}
and similarly for negatively oriented biased agents:
\begin{equation*}
\hat{x}_{\mathcal{B}^-}(t+1) = \frac{1}{k+1} \left [ k(1-f)(1-q)\hat{x}_{\mathcal{U}}(t) + (1-q)kf^+(t)\hat{x}_{\mathcal{B}^+}(t) + ((1-q)kf^-+1)(t)\hat{x}_{\mathcal{B}^-}(t) \right ] \ .
\end{equation*}
	
We have therefore established the update rule for the expected signal mix of the three sub-populations at any time $t$. We collate this update rule into a matrix form for convenience:
\begin{equation} \label{eq:xidyn}
\boldsymbol{\xi}(t+1) = \frac{1}{k+1}(\boldsymbol{F}(t)+\boldsymbol{I}_3) \boldsymbol{\xi}(t) + \boldsymbol{b} \ ,
\end{equation}
where
\begin{equation*}
\boldsymbol{\xi}(t) = [\hat{x}_{\mathcal{U}}(t), \hat{x}_{\mathcal{B}^+}(t), \hat{x}_{\mathcal{B}^-}(t)]^T \ , \qquad \boldsymbol{b} = \left [0, \frac{kq}{k+1}, 0 \right]^T
\end{equation*}
and
\begin{equation*}
\boldsymbol{F}(t)=k
\begin{bmatrix}
(1-f) & f^+(t) & f^-(t) \\
(1-q)(1-f) & (1-q)f^+(t) & (1-q)f^-(t) \\
(1-q)(1-f) & (1-q)f^+(t) & (1-q)f^-(t)
\end{bmatrix} .
\end{equation*}

If we further simplify notation by defining $\hat{\boldsymbol{F}}(t) = (\boldsymbol{F}(t)+\boldsymbol{I}_3)/(k+1)$, we get to the following compact expression for \eqref{eq:xidyn}:
\begin{equation*}
\boldsymbol{\xi}(t+1) = \hat{\boldsymbol{F}}(t) \boldsymbol{\xi}(t+1) + \boldsymbol{b} \ .
\end{equation*}

The long-run evolution of the signal mixes can be determined from the above equation if the evolution of $f^+(t)$ (and, consequently, of $f^-(t)$) in the matrix $\hat{\boldsymbol{F}}(t)$ is known. Assume for the moment we are at some time $t^*$ at which the system has settled, i.e., biased agents will keep their orientations intact and therefore will not cause the value of $f^+(t)$ to change for $t > t^*$. In this case we can write:
\begin{equation*}
\lim_{t \rightarrow \infty}\boldsymbol{\xi}(t) = \lim_{t \rightarrow \infty} \hat{\boldsymbol{F}}(t^*)^t \boldsymbol{\xi}(t^*) + (\hat{\boldsymbol{F}}(t^*)-\boldsymbol{I}_3)^{-1}\boldsymbol{b} \ .
\end{equation*}
It can be shown easily that, due to its double substochasticity, we have $\lim_{t \rightarrow \infty}\hat{\boldsymbol{F}}(t^*)^t = 0$, and therefore
\begin{equation*}
\lim_{t \rightarrow \infty}\boldsymbol{\xi}(t) = (\hat{\boldsymbol{F}}(t^*)-\boldsymbol{I}_3)^{-1}\boldsymbol{b} \ .
\end{equation*}
	
The above limit allows to calculate the steady state signal mixes for all sub-populations explicitly:	
		
\begin{equation} \label{eq:SSE}
\lim_{t \rightarrow \infty}\boldsymbol{\xi}(t) = 
\begin{bmatrix}
\hat{x}_{\mathcal{U}}^* \\
\hat{x}_{\mathcal{B}^+}^* \\
\hat{x}_{\mathcal{B}^-}^*
\end{bmatrix} =	
\begin{bmatrix}
f^+(t^*) / f\\
(1-q) f^+(t^*) / f + q\\
(1-q) f^+(t^*) / f
\end{bmatrix} = 
\begin{bmatrix}
\hat{y}_{\mathcal{B}}(t^*)\\
(1-q)\hat{y}_{\mathcal{B}}(t^*) + q\\
(1-q)\hat{y}_{\mathcal{B}}(t^*)
\end{bmatrix} .
\end{equation}
In the next Section we will approximate this result to the case where biased agents have not settled their orientation yet.

\subsection{Predicting the trajectory of biased agents' orientations}
	
Biased agents change their orientation when they receive a stream of incongruent signals that overcome their ability to distort them using confirmation bias. There are two points in the evolution of the model where this is possible. Firstly, this may happen in the early stages of the evolution, where the information sets held by the agents are relatively small and the stochasticity of the model can induce changes in orientation. Secondly, this may happen in the long run, where sustained changes in orientation can be brought along when one of the two camps of biased agents becomes able to systematically bias the available information. This leads to the composition of signals experienced by each node to change consistently in one direction, which can cause large scale switches in orientation, which in turn triggers a domino effect, as newly switched nodes will accelerate the rate at which signals are distorted.

Let us capture this notion more formally. Consider the expected long term signal mix of each sub-population assuming the biased agents have settled (\eqref{eq:SSE}). Suppose the positively oriented biased agents have an expected steady state signal mix $\hat{x}^*_{\mathcal{B}^+} < 1/2$. If such steady state value is to be reached, then some positively oriented biased agents' signal mixes must fall below $1 / 2$, thereby switching orientation to negative. If this happens, then $\hat{y}_{\mathcal{B}}(t)$ falls and the steady state signal mix for \textit{all} agents strictly decreases\footnote{This can be proven rigorously with the results from the previous Section: the steady state mix is $(I-Q)^{-1}R^+$. $R^+$ is a vector with $0$ for each negative biased agent, and $(I-Q)^{-1}$ is element-wise $> 0$. A biased agent switching to positive turns a previous zero element of $R^+$ to positive, and adds another strictly positive vector to the steady state signal mix. The same argument is made in reverse for a positive to negative switch}. This, in turn, means more positively oriented biased agents switch orientation to reach their steady state, and so forth until all such agents switch to a negative orientation, yielding $\hat{y}_{\mathcal{B}} = 0$. A corresponding outcome can be determined for negatively oriented biased agents all being converted. We can therefore determine, for any given $t^*$, the approximate conditions under which we expect all positively oriented biased agents to switch their orientations to negative in the eventual steady state. Setting $\hat{x}^*_{\mathcal{B}^+} < 1/2$ in \eqref{eq:SSE} we have
\begin{equation} \label{eq:low_f}
(1-q)\hat{y}_{\mathcal{B}}(t^*) + q < \frac{1}{2} \qquad
\Longrightarrow \qquad \hat{y}_{\mathcal{B}}(t^*) < \frac{1}{2-2q} \ ,
\end{equation}
and, correspondingly, for all negatively oriented biased agents to be tipped to positive we have the following condition:
\begin{equation} \label{eq:high_f}
\hat{y}_{\mathcal{B}}(t^*) > \frac{1-2q}{2-2q} \ .
\end{equation}
	
Let us now consider the case $t_0 = 0$, which means we are approximating the expected trajectory of the entire system given a starting fraction of positively oriented biased agents $f^{+*}(0) / f = \hat{y}_{\mathcal{B}}(0)$. We then have the following approximate result for the steady state signal mix of the \emph{unbiased} sub-population:

\begin{equation*} \label{eq:curtainplotsi}
\hat{x}^*_{\mathcal{U}} = \hat{y}^*_{\mathcal{B}} \approx \left\{
\begin{array}{ll}
\hat{y}_{\mathcal{B}}(0) & \text{for} \ \frac{1}{2(1-q)} \leq \hat{y}_{\mathcal{B}}(0) \leq \frac{1-2q}{2(1-q)} \\
1 & \text{for} \ \hat{y}_{\mathcal{B}}(0) > \frac{1-2q}{2(1-q)} \\
0 & \text{for} \  \hat{y}_{\mathcal{B}}(0) < \frac{1}{2(1-q)} \ , \\ 
\end{array}
\right.
\end{equation*}
where the latter two conditions derive from Eqs. \eqref{eq:low_f} and \eqref{eq:high_f}, while the first condition is the same reported in \eqref{eq:SSE} adapted for the case $t_0 = 0$.

\subsection{Steady state signal mix variance.}

In the previous Section we have provided approximations for the first moment of the 	steady state signal mixes of the unbiased agents, as well as those of the two biased agent sub-populations. We have also approximated the long term ``settled'' fractions of positively and negatively oriented biased agents. We noted that for $\hat{y}_{\mathcal{B}}(0) > \frac{1-2q}{2(1-q)}$ ($\hat{y}_{\mathcal{B}}(0) < \frac{1}{2(1-q)}$), the steady state signal mix is likely to asymptotically reach $1$ ($0$). Under these conditions, all agents eventually trivially possess the same signal $+1$ ($-1$). Therefore the distribution of signal mixes tends asymptotically to a Dirac distribution on $1$ ($0$).

We therefore proceed with the assumption that $\frac{1}{2(1-q)} \leq \hat{y}_{\mathcal{B}}(0) \leq \frac{1-2q}{2(1-q)}$, and as such use (\ref{eq:SSE}) to approximate:

\begin{equation*} 
\lim_{t \rightarrow \infty}\boldsymbol{\xi}(t) = 
\begin{bmatrix}
\hat{y}_{\mathcal{B}}(t^*)\\
(1-q)\hat{y}_{\mathcal{B}}(t^*) + q\\
(1-q)\hat{y}_{\mathcal{B}}(t^*)
\end{bmatrix} \approx
\begin{bmatrix}
\hat{y}_{\mathcal{B}}(0)\\
(1-q)\hat{y}_{\mathcal{B}}(0) + q\\
(1-q)\hat{y}_{\mathcal{B}}(0)
\end{bmatrix}
\end{equation*}

Correspondingly, we note $f^{+*} = \hat{y}_{\mathcal{B}}(t^*) = \hat{y}_{\mathcal{B}}(0)$. We would now like to characterise the distribution of signal mixes for each sub-population at the steady state beyond its first moment. We begin with an approximation of the variance, under the asymptotic limit of large populations $n \to \infty$.
	
For convenience we define the steady state signal mix variance of any sub-population $G$ as $\sigma^2_{G}$. In the following, we will provide approximate expressions for the steady state signal mix variances $\sigma^2_{\mathcal{U}}$, $\sigma^2_{\mathcal{B}^+}$, and $\sigma^2_{\mathcal{B}^-}$, and for the overall variance $\sigma^2_{V}$.
	
Consider an agent $i$ picked uniformly at random from the entire population. The variance of such an agent's steady state signal mix $\mathrm{Var}[\hat{x}_i^*]$ represents the variance across the entire population $\sigma^2_V$. From the law of total variance, this can be broken down as follows
\begin{equation} \label{eq:LTV}
\sigma^2_{V} = \mathrm{Var}[\hat{x}_i^*] = \mathbb{E} \big [ \mathrm{Var}[\hat{x}_i^*] \ | \ i \in \{\mathcal{U}, \mathcal{B}^+, \mathcal{B}^-\} \big ] + \mathrm{Var} \big [ \mathbb{E} \left [ \hat{x}^*_i \ | \ i \in \{U, \mathcal{B}^+, \mathcal{B}^-\} \right ]  \big ] \ ,
\end{equation}
where
\begin{equation} \label{eq:CondVar}
\mathbb{E} \big [ \mathrm{Var}[\hat{x}_i^*] \ | \ i \in \{\mathcal{U}, \mathcal{B}^+, \mathcal{B}^-\} \big ] = (1-f) \sigma^2_{\mathcal{U}} + f^{+*} \sigma^2_{\mathcal{B}^+} + f^{-*} \sigma^2_{\mathcal{B}^-} \ ,
\end{equation}
and	
\begin{eqnarray} \label{eq:var2}
\mathrm{Var} \big [ \mathbb{E} \left [ \hat{x}^*_i \ | \ i \in \{\mathcal{U}, \mathcal{B}^+, \mathcal{B}^-\} \right ]  \big ] &=&
(1-f) \bigg \{ \frac{f^{+*}}{f} - \mathbb{E} \Big [ \mathbb{E} [ \hat{x}^*_i \ | \ i \in \{\mathcal{U}, \mathcal{B}^+, \mathcal{B}^-\}] \Big ] \bigg \}^2 \\ \nonumber
&+& f^{+*} \bigg \{ (1-q)\frac{f^{+*}}{f}+q - \mathbb{E} \Big [ \mathbb{E} [ \hat{x}^*_i \ | \ i \in \{\mathcal{U}, \mathcal{B}^+, \mathcal{B}^-\}] \Big ] \bigg \}^2 \\ \nonumber
&+& f^{-*} \bigg \{ (1-q)\frac{f^{+*}}{f} - \mathbb{E} \Big [ \mathbb{E} [ \hat{x}^*_i \ | \ i \in \{\mathcal{U}, \mathcal{B}^+, \mathcal{B}^-\}] \Big ] \bigg \}^2 \ .
\end{eqnarray}
Noting that
\begin{equation*}
\mathbb{E} \Big [ \mathbb{E} [ \hat{x}^*_i \ | \ i \in \{\mathcal{U}, \mathcal{B}^+, \mathcal{B}^-\}] \Big ] = (1-f) \frac{f^{+*}}{f}+ f^{+*} \left ((1-q)\frac{f^{+*}}{f}+q \right) + f^{-*} \left ((1-q)\frac{f^{+*}}{f} \right) = \frac{f^{+*}}{f}
\end{equation*}
we can considerably simplify \eqref{eq:var2}:	
\begin{equation*}
\mathrm{Var} \big [ \mathbb{E} \left [ \hat{x}^*_i \ | \ i \in \{\mathcal{U}, \mathcal{B}^+, \mathcal{B}^-\} \right ]  \big ] = \frac{q^2 f^{+*} f^{-*}}{f} = f q^2 \hat{x}^*_{\mathcal{U}} (1-\hat{x}^*_{\mathcal{U}}) \ ,
\end{equation*}
where in the last step we have used the fact that $\hat{x}^*_{\mathcal{U}} = f^{+*} / f$, as per \eqref{eq:SSE}.
	
\eqref{eq:var2} provides a compact expression for the second contribution for the overall variance $\sigma^2_V$ in \eqref{eq:LTV}. We now turn to the first term (\eqref{eq:CondVar}). In order to be able to calculate it, we must compute the variance of each sub-population. Let us begin with the unbiased agent sub-population:

\begin{equation} \label{eq:sigmaU}
\sigma^2_{\mathcal{U}} = \mathrm{Var}[\hat{x}^*_{\mathcal{U}}] = \mathrm{Var} \left [ \frac{1}{k}\sum_{j \in \partial_i} \hat{x}^*_j \right ] = 
\frac{\sigma^2_{V}}{k} + \frac{2}{k^2}\sum_{(j,\ell) \in \partial_i} \mathrm{Cov} [\hat{x}^*_j,\hat{x}^*_\ell] = \frac{\sigma^2_{V}}{k} + \mathcal{O} \left (\frac{1}{k^2} \right) \ ,
\end{equation}
where in the last term we have assumed the covariance term to decay as $k^{-2}$, which will be proved in the next Section. In analogy with the above results, for the biased agent sub-population (either positively or negatively oriented) we have:
\begin{equation} \label{eq:sigmaB}
\sigma^2_{\mathcal{B}} = \mathrm{Var} [\hat{x}^*_{i_B}] =  \frac{(1-q)^2 \sigma^2_{V}}{k}  + \mathcal{O} \left (\frac{1}{k^2} \right) \ .
\end{equation}

Substituting the two expressions above in \eqref{eq:CondVar}, and combining the result with the one obtained in \eqref{eq:var2}, we finally obtain the following result for the overall variance in \eqref{eq:LTV}:
\begin{equation*}
\sigma^2_{V} =  \frac{1-fq(2-q)}{k} \sigma^2_{V}  + f q^2 \hat{x}^*_{\mathcal{U}} (1-\hat{x}^*_{\mathcal{U}}) + \mathcal{O}\left(\frac{1}{k^2}\right) \ .
\end{equation*}
Solving for $\sigma^2_{V}$ we get
\begin{equation*}
\sigma^2_{V} = \frac{k f q^2 \hat{x}^*_{\mathcal{U}} (1-\hat{x}^*_{\mathcal{U}})}{k+fq(2-q)-1} + \frac{k}{k+fq(2-q)-1} \mathcal{O}\left(\frac{1}{k^2}\right) \approx  f q^2 \hat{x}^*_{\mathcal{U}} (1-\hat{x}^*_{\mathcal{U}})  \ ,
\end{equation*}
where we have used the fact that $fq(2-q)-1 \in [-1,0]$, and therefore $fq(2-q)-1 \ll k$ even for moderate connectivity.   

Finally, we can specialize the above result to the three sub-populations via Eqs. \eqref{eq:sigmaU} and \eqref{eq:sigmaB}:
\begin{eqnarray} \label{eq:variances}
\sigma^2_{\mathcal{U}} &\approx& \frac{f q^2 \hat{x}^*_{\mathcal{U}} (1-\hat{x}^*_{\mathcal{U}})}{k}\\
\sigma^2_{\mathcal{B}_\pm} &\approx&  \frac{f \left(q(1-q) \right)^2 \hat{x}^*_{\mathcal{U}} (1-\hat{x}^*_{\mathcal{U}})}{k} \ .
\end{eqnarray}	
	


\subsection{Explicit neighbourhood covariance expressions.}
	
In this Section we establish that the covariance term appearing in \eqref{eq:sigmaU} can indeed be assumed to be of order $k^{-2}$. The first thing to note is that the term $\mathrm{Cov}[\hat{x}^*_j,\hat{x}^*_\ell]$ for two generic agents can be bounded above by the covariance $\mathrm{Cov}[\hat{x}^*_j,\hat{x}^*_\ell | j,\ell \in U]$ between unbiased agents' steady state signal mixes. To see this, suppose $j,\ell \in \mathcal{B}^+$:
\begin{eqnarray*}
\mathrm{Cov}[\hat{x}^*_j,\hat{x}^*_\ell | j,\ell \in \mathcal{B}^+] &=& \mathrm{Cov} \left [ \frac{(1-q)\sum_{h \in \partial_j} \hat{x}^*_h + qk}{k},\frac{(1-q)\sum_{m \in \partial_\ell} \hat{x}^*_m + qk}{k} \right ] \\ \nonumber
&=& \frac{(1-q)^2}{k^2} \mathrm{Cov} \left[\sum_{h \in \partial_j} \hat{x}^*_h,\sum_{m \in \partial_\ell} \hat{x}^*_m \right] 
< \frac{1}{k^2} \mathrm{Cov} \left [ \sum_{h \in \partial_j} \hat{x}^*_h,\sum_{m \in \partial_\ell} \hat{x}^*_m \right ] \\ \nonumber
&=& \mathrm{Cov}[\hat{x}^*_j,\hat{x}^*_\ell | j,\ell \in U] \ .
\end{eqnarray*}
\begin{figure}[h!]
\centering
		\includegraphics[width=0.3\textwidth]{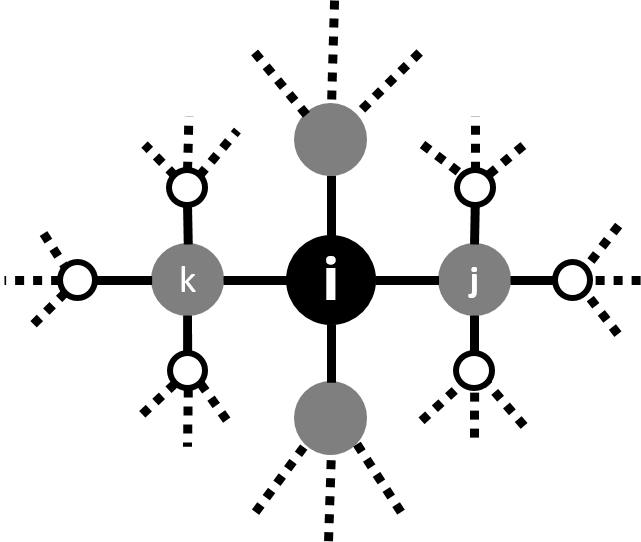}
	\caption{The covariance of $i$'s neighbours, $j$ and $k$ (grey), can be decomposed into the covariance between its neighbours (black and white). This consists of the covariance between neighbours that are two steps apart (black to white) as well as those that have distance four steps apart (white to white).}
	\label{fig:CovTerms}
\end{figure}
Let $\mathrm{Cov}(\hat{x}^*_j,\hat{x}^*_l|(j,l) \in U) = \sigma_2$, denote the least upper bound for the covariance between two nodes of distance $2$ apart. Similarly, let $\sigma_d$ be the same for nodes of distance $d$ apart. In the remainder of this section we are seeking to establish a relationship between these \textit{upper bounds} and in doing so recursively determine the upper bound at $d=2$.

It is worth reiterating that we are approximating the variance of the steady state signal mixes at the asymptotic limit $n \to \infty$. Given this assumption, a $k$-regular tree will approximate a Cayley tree. A useful consequence of this assumption is that the network is globally tree-like, and loops vanish in the limit. As such, only a single path exists between any two nodes.
Therefore, in \eqref{eq:sigmaU} we have:
\begin{equation*}
\sigma^2_{\mathcal{U}} =  \frac{\sigma^2_V}{k} + \frac{2}{k^2}\sum_{(j,\ell) \in \partial_i}\mathrm{Cov}[\hat{x}^*_j,\hat{x}^*_\ell] \leq \frac{\sigma^2_V}{k} +\frac{2}{k^2}\sum_{(j,\ell) \in \partial_i} \sigma_2 = \frac{\sigma^2_V}{k} +\frac{\sigma_2}{k}(k-1) \ .
\end{equation*}
Therefore we can establish that if $\sigma_2 = \mathcal{O}(k^{-2})$, the whole expression will be of order $\mathcal{O}(k^{-2})$. To do this note:
\begin{eqnarray*}
\mathrm{Cov}[\hat{x}_j^*, \hat{x}_\ell^*] &=& \frac{1}{k^2} \mathrm{Cov}\left [\hat{x}_i^* + \sum_{m \in \partial_j/i}\hat{x}_m^* , \hat{x}_i^* + \sum_{n \in \partial_\ell/i}\hat{x}_n^*\right ] = \frac{1}{k^2} \left (\mathrm{Cov}[\hat{x}_i^*,\hat{x}_i^*]+  \sum_{(m,n) \in [\partial_j\times \partial_\ell] / (i,i)} \mathrm{Cov}[\hat{x}_m^*,\hat{x}_n^*] \right) \\ \nonumber
&=& \frac{1}{k^2} \left ( \mathrm{Cov}[\hat{x}_i^*,\hat{x}_i^*] + \sum_{m \in \partial_j/i} \mathrm{Cov}[\hat{x}_i^*,\hat{x}_m^*] +  \sum_{n \in \partial_\ell/i} \mathrm{Cov}[\hat{x}_i^*,\hat{x}_n^*] + \sum_{(m,n) \in [\partial_j/i \times \partial_\ell/i]} \mathrm{Cov}[\hat{x}_m^*,\hat{x}_n^*] \right ) \ .
\end{eqnarray*}
In this last step, we explicitly break down the covariance sum into the covariance between the neighbours of $j$ and $l$, which exist at various distances to one another. This is illustrated in \ref{fig:CovTerms}.

We can group these covariance pairs by their distance and bound them using our defined bounds $\sigma_d$:

\begin{equation*}
\mathrm{Cov}[\hat{x}_j^*, \hat{x}_\ell^*] \leq \frac{1}{k^2} \left ( \sigma^2_{\mathcal{U}}+ 2(k-1)\sigma_2 + (k-1)^2 \sigma_4 \right ) \ .
\end{equation*}
This allows us to recursively define
\begin{equation*}
\sigma_2 = \frac{1}{k^2} \left (\sigma^2_{\mathcal{U}}+ 2(k-1)\sigma_2 + (k-1)^2 \sigma_4 \right ) \ .
\end{equation*}
Re-arranging this expression we get:
\begin{equation*}
\sigma_2 = \frac{1}{(k-1)^2+1} \sigma^2_{\mathcal{U}}+ \frac{(k-1)^2}{(k-1)^2-1} \sigma_4 = \frac{1}{(k-1)^2+1} \sigma_0+ \frac{(k-1)^2}{(k-1)^2-1} \sigma_4 \ .
\end{equation*}
In the final step, we have replaced $\sigma^2_{\mathcal{U}}$ with $\sigma_0$ - which emphasizes that this term is merely the covariance of a node with a node at distance $0$ (i.e. its own variance), and $\sigma^2_{\mathcal{U}}$ is the largest possible variance expression amongst the biased and unbiased nodes. We can easily (though quite tediously) repeat this process for the covariance of nodes at any distance $d$ to establish:
\begin{equation*}
\sigma_d = \frac{1}{(k-1)^2+1} \sigma_{d-2}+ \frac{(k-1)^2}{(k-1)^2-1} \sigma_{d+2} \ .
\end{equation*}
This linear recurrence relation can be solved with the boundary conditions that $\sigma_0 = \sigma^2_{\mathcal{U}}$ and $\lim_{d \to \infty} \sigma_d = 0$\footnote{In other words, nodes at infinitely long distances have a covariance that decays to zero}. This establishes:
\begin{equation*}
\sigma_d = \frac{\sigma_{\mathcal{U}}^2}{(k-1)^d} \ ,
\end{equation*}
and therefore:
\begin{equation*}
\sigma_2 = \frac{\sigma_{\mathcal{U}}^2}{(k-1)^2} = \mathcal{O}(k^{-2}) \ .
\end{equation*}
Therefore, to finalise \eqref{eq:sigmaU}:
\begin{equation*}
\sigma^2_{\mathcal{U}} =  \frac{1}{k}\sigma^2_V + \mathcal{O} \left(\frac{1}{k^2}(k^2-k)\sigma_2\right) = \frac{1}{k}\sigma^2_V+\mathcal{O}(k^{-2}) \ .
\end{equation*}
	
\subsection{Steady state signal mix normality.}
\label{SSBN}
	
We finally proceed to demonstrate that the distribution of signal mixes is approximately normal when $k$ and $n$ are large. Assume firstly that $n \rightarrow \infty$, which ensures that the model's $k$-regular network becomes a Cayley tree with no loops. Let us also assume that $k = \epsilon n$ for some $\epsilon > 0$, ensuring $k$ also grows arbitrarily large, but still can be arbitrarily smaller than $n$.

As we have already established in the previous section, the covariance between the steady state signal mixes of unbiased agents at distance $2$ decays as $k^{-2}$, which implies that such signals mixes become asymptotically independent in the aforementioned limits. Therefore, the steady state signal mix of an unbiased agent $\hat{x}_{i_{\mathcal{U}}}^* = \sum_{j \in \partial_i} \hat{x}^*_j / k$ becomes the sum of an infinitely large set of independent random variables. Furthermore, the variables will follow one of three distributions, depending on which sub-population the agent's neighbors belong to:
	
\begin{equation*}
\hat{x}_{i_{\mathcal{U}}}^*=\frac{1}{k}\sum_{j\in \partial_i} \hat{x}_j^* =\frac{1}{k} \left (\sum_{j\in \partial_i \cap U} \hat{x}_j^*  + \sum_{j\in \partial_i \cap \mathcal{B}^+} \hat{x}_j^* + \sum_{j\in \partial_i \cap \mathcal{B}^-} \hat{x}_j^* \right ) \ .
\end{equation*}
Each of the above contributions is a sum of an infinitely large set of independent and identically distributed variables, which implies that each of them is normally distributed. This, in turn, implies that the steady state signal mixes of the unbiased agents (and, by generalisation, of the biased agents) is asymptotically normal. From the results obtained for the first two moments of the signal mix distributions in the previous sections (see Eqs. \eqref{eq:SSE} and \eqref{eq:variances}), we can conclude that when $n \rightarrow \infty$ and $k = \epsilon n$ we have
\begin{eqnarray*} \label{SSNormal}
\hat{x}_{\mathcal{U}} &\overset{d}{\to}& \mathcal{N} \left (\hat{x}^*_{\mathcal{U}},\frac{f q^2 \hat{x}^*_{\mathcal{U}}(1-\hat{x}^*_{\mathcal{U}})}{k} \right) \\ \nonumber
\hat{x}_{\mathcal{B}^+} &\overset{d}{\to}& \mathcal{N} \left((1-q)\hat{x}^*_{\mathcal{U}} + q,  \frac{f \left(q(1-q) \right)^2 \hat{x}^*_{\mathcal{U}} (1-\hat{x}^*_{\mathcal{U}})}{k} \right) \\ \nonumber
\hat{x}_{\mathcal{B}^-} &\overset{d}{\to}& \mathcal{N} \left ((1-q)\hat{x}^*_{\mathcal{U}} , \frac{f \left(q(1-q) \right)^2 \hat{x}^*_{\mathcal{U}} (1-\hat{x}^*_{\mathcal{U}})}{k} \right) \ .
\end{eqnarray*}

The normality of the distribution for the unbiased agents is demonstrated in Figure \ref{fig:fktrade}.	

\section{Section S5: Accuracy}

We can now aggregate our results in order to approximate the accuracy of a social network. As described in the main body of the paper, the accuracy of a network $\mathcal{A}(G)$ is the expected fraction of accurate unbiased agents in the steady state, i.e. accuracy quantifies the probability $\mathrm{Prob}(y_{i_{\mathcal{U}}}=+1)$ that a randomly picked unbiased agent in a random realisation of the model will correctly learn the ground truth\footnote{The definition of accuracy could very easily be extended to all agents instead of just unbiased agents, but we retain discussion to unbiased agents for simplicity}.
	
This is of course a complex outcome determined by a dynamic series of processes worth recapping. First, the model will generate initial signals for all agents, both biased and unbiased. All agents will share their signals, but biased agents will selectively sample incoming signals based on their current orientation. Over time, biased agents are able to influence the set of signals in the system, and the system converges towards a steady state where each agent possesses an equilibrium mix of signals. Accurate agents are those whose equilibrium signal mix is contains more positive than negative signals, i.e. $x_i^* > 1 / 2$.
	
In the previous Sections we have calculated the distribution of the initial signals, as well as the approximate steady state signal mix for a given set of initial signals (see Equation \ref{eq:curtainplotsi}). We have also approximated the steady state individual signal mix distributions (see Eq. \ref{SSNormal}), and as such we can approximate the fraction of accurate unbiased agents for a given steady state, which reads
\begin{equation} \label{eq:acc}
A(\mathcal{G}) = \frac{1}{2} \int_{0}^{1} dx_{\mathcal{U}}^* \ P(x_{\mathcal{U}}^*) \ \mathrm{erfc} \left (\frac{\frac{1}{2}-x_{\mathcal{U}}^*}{\sqrt{2} \ \sigma_{\mathcal{U}}} \right) \ ,
\end{equation}
where $P(x^*_{\mathcal{U}})$ is the distribution of the average signal mix across unbiased agents as determined by \eqref{eq:curtainplotsi} (see Eq. (3) of the main paper), while the complementary error function quantifies the fraction of unbiased agents whose steady state signal mix is above $1/2$, and are therefore accurate, under the normal approximation outlined in the previous section.
	
\begin{figure}[h!]
\includegraphics[width=10cm]{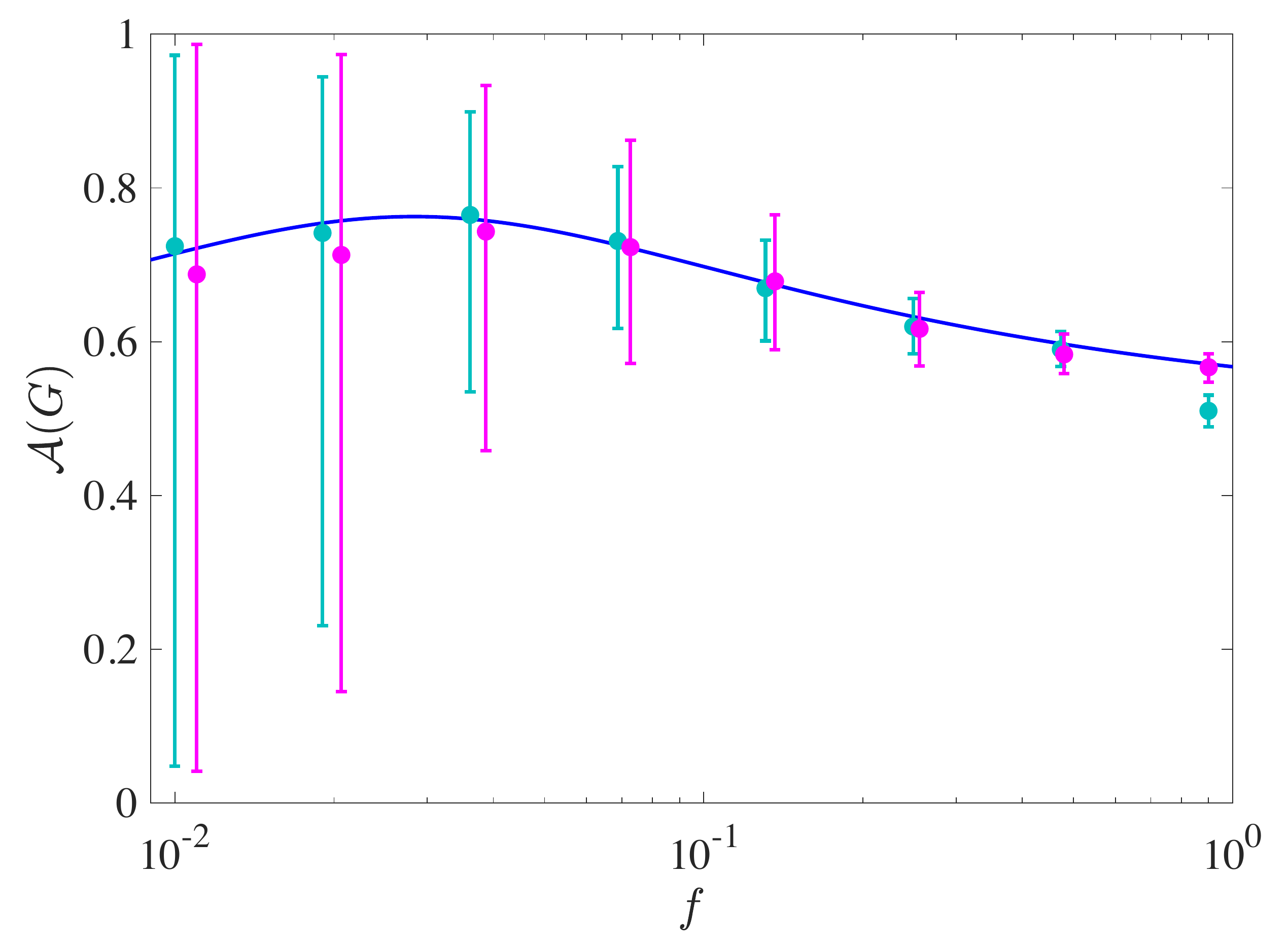}
\centering
\caption{Non-monotonic changes in expected accuracy as $f$ increases. The model's prediction are compared to numerical results obtained with simulations on both $k$-regular (light blue) and Erd\H os-R\'enyi (purple) networks. The parameters used in the simulations were $n = 10^4$, $p = 0.53$, $k = 8$, $q = 1$.}
\label{fig:accuracysi}
\end{figure}

Both ingredients employed in \eqref{eq:acc} have been obtained based on a number of approximations and asymptotic assumptions. We checked how such approximations hold against numerical simulations of the model's dynamics. The results are shown in Fig. \ref{fig:accuracysi} both for $k$-regular and Erd\H os-R\'enyi networks. As can be seen, the average accuracy obtained across independent numerical simulations of the model closely matches the expected value obtained with \eqref{eq:acc}, even for relatively low network size and average degree (the results reported were obtained for $n = 10^4$ and $k = 8$). The wider error bars for lower $f$ reflect the expected outcome that most runs of the model will result in total consensus on either $X = \pm 1$ (and therefore all accurate or all inaccurate agents), whereas as $f$ grows, the agents are highly polarised and the fraction of accurate and inaccurate agents will be relatively constant. 

\section{Section S6: Regression results}

\subsection{Theory and model interpretation.}
	
Our model is stylized, and therefore largely agnostic as to a particular interpretation of its parameters. Nevertheless, it is quite well suited to provide an initial exploration on a number of issue. In this Section, we shall test the model's ability to shed light on the impact that Internet access has on shaping popular opinion on specific issues (global warming in this case). In order to do this, we first specify how we are going to relate our model's parameters to real-world measurable quantities. 
	
	There are two convenient (and pragmatically equivalent) interpretations of the model in the context of Internet use. Consider the agent-specific ghost node interpretation, where each ghost node attached to a biased agent represents an aggregation of the ``filter bubble" (passive algorithmic affects) and ``selective exposure" (actively selecting information in a biased way) effects. An increase in Internet access therefore translates to an increase in access to these self-confirmatory effects, and corresponds to changing unbiased agents into biased agents (an increase in $f$). Alternatively, one could consider a scenario where the fraction of biased agents is fixed, in which case an increase in Internet would improve their ability to obtain self-confirmatory information (an increase in $q$). For the purposes of this exploration, however, the two effects are equivalent, and for convenience we only retain the interpretation where $f$ increases.
	
	As far as the interpretation of the degree variable $k$ is concerned, the important distinction to make here is that we are not interested in ``social networks'' as a catch-all term for the number of family and friends one has. Rather, given the model, we are interested in the degree to which individuals actively exchange information with their underlying social network with regards to the topic of interest. Therefore, for $k$ we wish to measure the volume of active social information diffusion in a given population.
	
	As per \eqref{eq:variances}, one of our model's main results is that $f$ and $k$ work in opposite directions when it comes to polarisation\footnote{Strictly speaking the result refers to the variance in information sets, but we exploit the monotonic relationship between information variance $\sigma_{x^*_{\mathcal{U}}}^2$ and polarization $z_{\mathcal{U}}^*$ for the remainder of this section} - an increase in confirmatory behaviours increases polarisation and is equivalent to a reduction in social information. Furthermore, if the majority of the population accurately learns the ground truth ($x_{\mathcal{U}}^*>1/2$), reductions in polarisation can be translated to an increase in consensus on the truth, as a smaller fraction of the population will arrive at inaccurate beliefs.
	
	Translated to current research on the role of the Internet, we attempt to use our model to shed light on what has been thought of as the dichotomous effects of Internet access on social learning and polarisation. On one hand it has been argued that Internet access improves exposure to diverse information via social networks \cite{haythornthwaite,pew2,leekim}, whereas on another it has been argued that Internet access enables confirmation bias on a previously unprecedented scale \cite{lelkes2017hostile,del2016spreading}. These contradictory effects may be in part responsible for the range of conflicting results obtained in recent research, and in our closing remarks we revisit some of these existing results in the context of our model.

\subsection{Data sources and measuring variables}
	
In order to test the model's predictions in the aforementioned context, we gathered data from the Yale Programme on Climate Change Communication 2016 Opinion Maps \cite{howe}, which provides state and county level survey data on opinions on global warming, as well as behaviours such as the propensity to discuss climate change with friends and family. We combined this with FCC 2016 county level data on residential high speed Internet access \cite{FederalCommunicationsCommission}. Finally, we also used a supplemental source in the data aggregated by the Joint Economic Council's Social Capital Project \cite{jec}, a government initiative aiming to measure social capital at a county level by aggregating a combination of state and county level data from sources such as the American Community Survey, the Current Population Survey, and the IRS.
	
	In this context, we measured accuracy as the estimated fraction of the population believing that ``global warming is happening''. We refer to this as ``GW Accuracy''. In other words, we are attempting to examine the degree to which social information and access to confirmatory bias mechanisms affect the ability of individuals to accurately learn an objective, measurable and uncontroversial ground truth (that global temperatures are rising).
	
	Internet access is measured by the FCC's data on county-level high speed broadband penetration amongst residentials (in \cite{lelkes2017hostile}, the authors utilise another instrumental variable approach to argue that increased broadband penetration does in fact increase Internet use). In Table \ref{table:regression1} we demonstrate preliminary ordinary least squares regression results by regressing GW Accuracy on Internet access, accounting for a range of covariates such as median age, median income, county population size and the fraction of adults with college degrees. The results indicate that even after controlling for relevant covariates, the net effect of Internet access on accuracy is positive\footnote{One may note that the impact of median income on this regression, and all subsequent results. is negative. We have verified this result through a number of additional checks. It appears that the inclusion of college education heavily affects this coefficient, implying that the effect of income on global warming beliefs is heavily mediated by access to education. We also performed some further checks by including dummy variables for political partisanship using county level voting results for the 2016 presidential elections. While political partisanship provides additional explanatory power over and above the current set of variables, the coefficient for income when including it is still negative. Unpacking the exact nature of this relationship would require a broader range of economic and political factors, which is clearly outside the scope this initial analysis, so we exclude partisanship and continue with the original model, allowing the coefficients to be taken at face value.} (and by interpretation, the effect of polarisation on this particular ground truth is negative).
	
	However, this alone is insufficient as research indicates Internet access is likely to improve the degree to which individuals can communicate information to friends and family, which in our model is precisely the variable $k$. The Yale Climate Change data includes a measure estimating the fraction of the county population that discusses global warming regularly with family and friends (``Social Discussion''). To sense check this, Table \ref{table:regression1} (column 2) demonstrates that increased Internet access does indeed improve the ability to discuss matters with friends and family, even after controlling for relevant covariates, which is consistent with a broad set of empirical research on the topic (see \cite{wellman} for a review).
	
	Therefore, this allows us to construct our final model in Table 1(3) where we regress GW Accuracy on \textit{both} Social Discussion and Internet access (and the covariates). We can now interpret the coefficient on Internet access as the residual effect of Internet access \textit{after} controlling for the effect it has on Social Discussion. One way of thinking about this is to consider all causal pathways from Internet access to belief formation - some fraction of them will be via improved access to social and discussion networks (communication platforms, online social networks, and forums), and the remaining fraction will be non-social (algorithmic effects, filter bubbles, online news media, selective exposure, etc). By accounting for the former effects by observing the discussion network size in Social Discussion, the residual effect of internet access will aggregate all these other effects. This lines up with the interpretation of $f$ in our model - Internet users will have access to these effects (``biased agents'') and non-Internet users will not. The results confirm our hypothesis - Social Discussion ($k$) and residual Internet Access ($f$) act in opposite directions when it comes to learning the ground truth, even after conditioning on a range of covariates.
	
	It is worth unpacking these results in detail. The direct effect of a $1$ percentage point increase in Internet access on global warming accuracy is negative\footnote{Table 1, Column 3, Row 2.} ($-2.400$). The direct effect on social discussion is extremely positive\footnote{Table 1, Column 2, Row 2.} ($3.736$), which leads to a corresponding improvement in accuracy\footnote{Table 1, Column 3, Row 1.} of $1.057 \times 3.736 \approx 3.95$. The net effect, of course, is positive ($3.95 - 2.40 = 1.55$),	 as indicated in the original, simple regression\footnote{Table 1, Column 1, Row 2.}. However, breaking down the causal mechanism into its constituent elements - direct internet use effects vs socially mediated internet effects - allows us the capture the nuance of what is actually happening.

\linespread{0.9}

\begin{table}
\caption{Initial Regression Results} 
\begin{tabular}{lccc}
\toprule
& \multicolumn{3}{c}{\textit{Dependent variable:}} \\
\cline{2-4} 
& \multicolumn{1}{c}{GW Accuracy} & \multicolumn{1}{c}{Social Discussion} & \multicolumn{1}{c}{GW Accuracy} \\ 
& \multicolumn{1}{c}{(1)} & \multicolumn{1}{c}{(2)} & \multicolumn{1}{c}{(3)}\\ 
\midrule
Social Discussion &  &  & 1.057$^{***}$ \\ 
			&  &  & (0.019) \\ 
			& & & \\ 
Internet Access & 1.550$^{**}$ & 3.736$^{***}$ & -2.400$^{***}$ \\ 
			& (0.664) & (0.447) & (0.471) \\ 
			& & & \\ 
Median Age & -0.044$^{***}$ & -0.020$^{*}$ & -0.023$^{*}$ \\ 
			& (0.017) & (0.011) & (0.012) \\ 
			& & & \\ 
log(Median Household Income) & -6.691$^{***}$ & -1.659$^{***}$ & -4.938$^{***}$ \\ 
			& (0.464) & (0.313) & (0.327) \\ 
			& & & \\ 
log(Total Pop) & 0.690$^{***}$ & -0.399$^{***}$ & 1.112$^{***}$ \\ 
			& (0.071) & (0.048) & (0.050) \\ 
			& & & \\ 
College Education & 0.335$^{***}$ & 0.304$^{***}$ & 0.013 \\ 
			& (0.013) & (0.009) & (0.011) \\ 
			& & & \\ 
Constant & 123.417$^{***}$ & 43.730$^{***}$ & 77.178$^{***}$ \\ 
			& (4.846) & (3.267) & (3.501) \\ 
\midrule
Observations & \multicolumn{1}{c}{2,933} & \multicolumn{1}{c}{2,933} & \multicolumn{1}{c}{2,933} \\ 
$R^{2}$ & \multicolumn{1}{c}{0.312} & \multicolumn{1}{c}{0.448} & \multicolumn{1}{c}{0.662} \\ 
Adjusted $R^{2}$ & \multicolumn{1}{c}{0.311} & \multicolumn{1}{c}{0.447} & \multicolumn{1}{c}{0.661} \\ 
Residual Std. Error & \multicolumn{1}{c}{4.395 (df = 2927)} & \multicolumn{1}{c}{2.963 (df = 2927)} & \multicolumn{1}{c}{3.082 (df = 2926)} \\ 
F Statistic & \multicolumn{1}{c}{265.316$^{***}$} & \multicolumn{1}{c}{474.324$^{***}$} & \multicolumn{1}{c}{953.601$^{***}$} \\
 & \multicolumn{1}{c}{(df = 5; 2927)} & \multicolumn{1}{c}{(df = 5; 2927)} & \multicolumn{1}{c}{(df = 6; 2926)}  \\
\bottomrule
\textit{Note:}  & \multicolumn{2}{r}{$^{*}p<$0.1; $^{**}p<$0.05; $^{***}p<$0.01} \\ 
\end{tabular}
\label{table:regression1}
\end{table}

\linespread{1}

\subsection{Accounting for simultaneous causality.}
	
	A clear shortcoming of the above analysis is the fact that the variable ``Social Discussion'' is likely to have a reverse causal relationship with the outcome variable of ``GW Accuracy''. That is, the more likely individuals are to believe global warming is happening, the more likely they are to discuss this topic with friends and family.
	
	In order to account for this, we will take an instrumental variable approach. That is, we need some instrument that can account for independent variation in discussion with family and friends, which is otherwise unlikely to affect the belief in global warming. We note as before that $k$ can be interpreted as the fraction of the ``underlying social network'' that is activated to transmit social information related to the topic of global warming. We are therefore interested in a variable that can measure the pre-existing strength of these underlying social networks. To do so, we make use of the Social Capital Project, a government research programme by the Joint Economic Committee that attempts to measure Social Capital at a state and county level throughout the US. Social Capital as defined in this study (and numerous others\footnote{i.e. Putnam \cite{putnam2000bowling} (1995, p.19), ``...social capital refers to connections among individuals' social networks and the norms of reciprocity and trustworthiness that arise from them''.}) refers broadly to something ``related to social relationships, social networks, and civil society''. More specifically, it is measured with an intention to reflect communities with ``an abundance of close, supportive relationships'' \cite{jec}. 
	
	The index itself measures a spectrum of factors, and in particular a ``Community Health'' subindex. The subindex is calculated as the leading principal component across a variety of state and county-level measures of community engagement (where people ostensibly meet and socialise with friends and family), including religious congregations,  non-religious non-profit activities, public meeting attendance, working with neighbours to fix things, attending a meeting where politics was discussed, etc. This index is then validated by examining bivariate correlations with a battery of county level benchmarks and measures of social dysfunction.
	
	The strength of this instrument is established in Table \ref{tab:regression2} (column 1), where a first stage least squares regression is run to show that improvements in Community Health do translate to improved discussion with friends and family (controlling for covariates).
	
	The validity is established through a series of additional checks. Factors such as religious attendance, public meetings, etc. are unlikely to have a causal effect on people's beliefs about global warming independent of them being a medium to allow for social discussion of these topics. The only other reasonable and plausibly significant causal channel is if these factors are caused by or cause an increase membership in social groups (for instance, political parties) that are strongly associated with reduced belief in global warming. In particular, it is well-established that members of the Republican Party have a reduced belief in the existence of Global Warming \cite{pewgw}. To check this, we examined the bivariate correlation between Community Health and the percentage of GOP votes cast in the 2016 presidential election. The results were weak, with a correlation of only $0.14$, meaning only $1.8\%$ of the variation in the measures were explained by the relationship.
	
	Having established the strength and validity of the instrument, we demonstrate the results from the two stage least squares regression results in Table \ref{tab:regression2} (column 2). We can see the qualitative results of the simpler model have been preserved, with the effects predictably attenuated. However, the results are still significant, and corroborate our theory. After separating out the social and confirmatory effects of Internet access, we can see the impact on Accuracy (and Polarisation) both occur in the direction that we predict.
	
	Once again, let us unpack the results. The direct effect of a $1$ percentage point increase in internet access on global warming accuracy is negative\footnote{Table 2, Column 2, Row 3.} ($-1.712$). The direct effect on social discussion is extremely positive\footnote{Table 2, Column 1, Row 2.} ($3.143$), which leads to a corresponding improvement in accuracy\footnote{Table 2, Column 2, Row 2.} of $0.872 \times 3.143 \approx 2.74$. The net effect, of course, is positive ($2.74 - 1.71 = 1.03$). Once again, breaking down the causal mechanism into its constituent elements - direct internet use effects vs socially mediated internet effects - allows us the capture the nuance of what is actually happening.
	
	It appears, for the topic of global warming, the net impact of Internet access on social learning is positive. Increase in Internet access has a direct negative impact on learning (via $f$, or $q$). However, it leads to a significant positive impact on social discussion ($k$), and the net result of this is positive. This result remains robust even after controlling for a battery of relevant covariates.
	
	It should be emphasized that this result is merely an initial exploration of how our model can provide some testable predictions to empirical data, as opposed to a detailed effort to understand the effect of Internet access on global warming beliefs. Having said that, the initial results are encouraging, and we hope the clarity of the analytic results of our model pave the way for testing variations of the idea of biased information aggregation in a range of outcomes and settings.
	
\linespread{0.9}	
	\begin{table}[!htbp] 
	\centering 
		\caption{IV Regression Results} 
		\begin{tabular}{lccc}
	\toprule
			& \multicolumn{2}{c}{\textit{Dependent variable:}} \\ 
			\cline{2-3} 
			\\[-1.8ex] & \multicolumn{1}{c}{Social Discussion} & \multicolumn{1}{c}{GW Accuracy} \\ 
			\\[-1.8ex] & \multicolumn{1}{c}{\textit{OLS}} & \multicolumn{1}{c}{\textit{instrumental}} \\ 
			& \multicolumn{1}{c}{\textit{(First Stage LS)}} & \multicolumn{1}{c}{\textit{variable (2SLS)}} \\ 
			\\[-1.8ex] & \multicolumn{1}{c}{(1)} & \multicolumn{1}{c}{(2)}\\ 
			\hline \\[-1.8ex] 
			Community Health Index & 1.501$^{***}$ &  \\ 
			& (0.081) &  \\ 
			& & \\ 
			Social Discussion &  & 0.872$^{***}$ \\ 
			&  & (0.060) \\ 
			& & \\ 
			Internet Access & 3.143$^{***}$ & -1.712$^{***}$ \\ 
			& (0.424) & (0.523) \\ 
			& & \\ 
			log(Median Household Income) & -1.814$^{***}$ & -5.281$^{***}$ \\ 
			& (0.297) & (0.346) \\ 
			& & \\ 
			log(Total Pop) & 0.300$^{***}$ & 1.044$^{***}$ \\ 
			& (0.059) & (0.056) \\ 
			& & \\ 
			Median Age & -0.081$^{***}$ & -0.028$^{**}$ \\ 
			& (0.011) & (0.012) \\ 
			& & \\ 
			College Education & 0.245$^{***}$ & 0.070$^{***}$ \\ 
			& (0.009) & (0.021) \\ 
			& & \\ 
			Constant & 42.621$^{***}$ & 85.649$^{***}$ \\ 
			& (3.095) & (4.356) \\ 
			& & \\ 
			\hline \\[-1.8ex] 
			Observations & \multicolumn{1}{c}{2,932} & \multicolumn{1}{c}{2,932} \\ 
			$R^{2}$ & \multicolumn{1}{c}{0.506} & \multicolumn{1}{c}{0.651} \\ 
			Adjusted $R^{2}$ & \multicolumn{1}{c}{0.505} & \multicolumn{1}{c}{0.651} \\ 
			Residual Std. Error (df = 2925) & \multicolumn{1}{c}{2.803} & \multicolumn{1}{c}{3.129} \\ 
			F Statistic & \multicolumn{1}{c}{499.387$^{***}$ (df = 6; 2925)} &  \\ 
			\bottomrule
				\textit{Note:}  & \multicolumn{2}{r}{$^{*}p<$0.1; $^{**}p<$0.05; $^{***}p<$0.01} \\ 
		\end{tabular} 
		\label{tab:regression2} 
	\end{table} 

\linespread{1}	

\subsection{Making sense of broader empirical results.}
    
    We have seen so far that our model can help us decompose the effect of internet access on learning in the specific case of global warming facts. We now see if the model can help us better understand the seemingly conflicting findings we have found in existing research as indicated above. It should be said that the following interpretations are meant only to be indicative of how our model can help shape our theoretical understanding of empirical phenomena, rather than a detailed exploration of the specific empirical questions these papers explore.
    
    In \cite{boxell2017greater}, the authors argue that internet access has not had an effect on political polarisation because the demographic with the lowest increase in internet use - the elderly - has had the highest increase in political polarisation. However, it is also well established that older people have smaller network sizes than younger people \cite{cornwell} and growing evidence of demographic shifts suggest that older people are increasingly living alone \cite{Berry2013}. This translates to a direct fall in $k$ for such populations, and without a corresponding increase in $k$ provided by internet access, we would in fact expect to see higher polarisation in such a group.

    In \cite{lelkes2017hostile}, the authors argue that an increase in internet access  leads to an increase in political polarisation. Firstly, it is worth noting that the overall effect size is very small - increasing the number of broadband providers by $10\%$ increases political polarisation by $0.003$ points (on a scale between $0$ and $1$). This is consistent with notion that social connectivity will dampen the direct effect of biased media, and it is possible one could uncouple the effect of the internet on social connectivity as opposed to enabling confirmation bias with some proxy measure for social connectivity. What is also noteworthy is that the researchers included the level of ``political interest'' per county as a mediating variable in parts of the analysis. So for example, if we allow $f$ to represent the fraction of respondents in each county with such strong partisan interest, then $q$ could represent the level of bias these agents can display due to access to partisan media on the internet. Under this interpretation we can make sense of the interaction terms in the regression results - the effect of internet access on polarisation was considerably higher for counties where political interest is higher, which is exactly what we predict from the product ($fq^2$) in \eqref{eq:variances}.

    In \cite{barbera2014social}, the author argues that internet access leads to a decrease in political polarisation. This study looks solely at Twitter networks over time (but shows how they relate to political polarisation data offline). The author finds that more diverse Twitter networks lead to reduced polarisation over time. Again, our model predicts the following - since everyone is already on Twitter in this scenario, the fractions $f$ and $q$ are untouched. However, the author notes that more diverse networks are directly correlated with larger networks - a larger $k$. It follows therefore that these users with reduced polarisation experienced an increased $k$ without a corresponding change in $f$ or $q$, and the results follow.
    
    All in all, our biased learning model has proven to provide useful insight into a long-standing debate about an important empirical topic. We show that it allows us compress a large and complex set of causal mechanisms in the literature down to the effect of three terms of interest - the prevalence of biased agents ($f$), degree of bias ($q$), and social connectivity ($k$). In doing so, we were able to shed insights on the mechanisms at play when it came to internet access, and provide the beginnings of a more uniform understanding of what previously conflicting data has suggested to date.

	\bibliography{bibliography}

\end{document}